\documentclass{article}

\usepackage{arxiv}      

\usepackage[utf8]{inputenc}
\usepackage[T1]{fontenc}

\usepackage{amsmath, amssymb, amsfonts, amsthm}
\usepackage{mathtools}
\usepackage{bm}
\usepackage{mathrsfs}

\usepackage{graphicx}
\usepackage{subfig}
\usepackage{float}

\usepackage{booktabs}
\usepackage{multirow}
\usepackage{makecell}

\usepackage{algorithm}
\usepackage{algorithmicx}
\usepackage{algpseudocode}

\usepackage[numbers]{natbib}
\usepackage{doi}

\usepackage{hyperref}
\usepackage{url}

\usepackage{xcolor}
\usepackage{textcomp}
\usepackage{manyfoot}
\usepackage{microtype}
\usepackage{nicefrac}

\usepackage[title]{appendix}

\usepackage{lineno}

\usepackage[nolist]{acronym}

\usepackage{listings}

\title{Influence of Extruded Filament Shape on Buildability in 3D Concrete Printing: A Geometry-Informed Deep Learning--FEM Approach}

\author{
    Giacomo Rizzieri$^{1,*}$,
    Saif-Ur-Rehman$^{2}$,
    Jörg F. Unger$^{2}$,
    Annika Robens-Radermacher$^{2}$\\
    $^{1}$Department of Civil and Environmental Engineering, Politecnico di Milano\\
    Piazza Leonardo da Vinci, 32, 20133, Milano, Italy\\
    $^{2}$Bundesanstalt für Materialforschung und -prüfung (BAM)\\
    Unter den Eichen 87, 12205, Berlin, Germany\\
    $^{*}$\texttt{giacomo.rizzieri@polimi.it}
}

\begin{document}

\maketitle


\begin{abstract}
The geometric morphology of deposited filaments can significantly influence the structural performance and stability of 3D concrete-printed (3DCP) structures. However, most finite element (FEM)-based approaches for buildability assessment represent printed layers as simplified rectangles, potentially limiting predictive accuracy. This study proposes a geometry-informed modelling framework that integrates the deep-learning-based filament shape prediction tool \textit{ShapeGen3DCP} with a layer-activation FEM approach to investigate the effect of realistic filament geometries on buildability. The framework generates geometry-aware numerical models directly from material and process parameters, eliminating the need for experimental filament characterization or computationally intensive fluid-flow simulations. Validation against experimental data and a parametric study of rectilinear walls demonstrate that extrusion parameters and the resulting filament geometry can significantly influence buildability predictions. Realistic filament representations are particularly important for free-flow deposition, whereas layer-pressing strategies are less sensitive to geometric simplifications. Among the investigated representations, an elliptical approximation provides an effective balance between geometric fidelity and modelling simplicity. When rectangular representations are preferred to enable regular computational meshes for faster simulations, defining their dimensions based on volume conservation improves prediction reliability compared with calibrating them using either the maximum filament width or the interlayer contact width. Overall, the proposed methodology demonstrates the importance of incorporating filament geometry into 3DCP simulations and provides practical guidance for selecting efficient and accurate geometric representations for buildability assessment.
\end{abstract}



\section{Introduction}\label{sec:1}
3D Concrete Printing (3DCP) is an emerging additive manufacturing technology that involves the layer-by-layer extrusion of cementitious mortar to fabricate complex objects and structural components. It offers the potential to improve design flexibility, material efficiency, and construction automation by enabling geometrically optimized structures and reducing reliance on conventional formwork.

The geometric fidelity and surface quality of a printed element are strongly governed by the morphology of the deposited layers \cite{wolfs2021}. Beyond aesthetic considerations, layer geometry and interlayer morphology can significantly influence interlayer bond strength \cite{marchment2019}, the structural performance of the hardened material \cite{nerella2019}, and durability-related properties \cite{nodehi2022}, for instance by modifying the ingress pathways available to deleterious agents within the hardened material \cite{bran-anleu2023}. By contrast, comparatively little is known about the influence of filament shape on buildability, defined as the maximum number of layers that can be successfully stacked with a given material and set of process conditions before structural collapse occurs. In particular, it remains unclear whether, and to what extent, layer geometry contributes to the mechanical response and overall stability of the printed object during deposition.

This lack of quantitative characterization is also reflected in the geometrical simplifications adopted by solid-based numerical models for simulating buildability in 3D concrete printing. Over the years, numerous approaches have been proposed, ranging from early FE-based formulations employing layer-by-layer activation to evaluate buckling and plastic collapse in rectilinear and cylindrical structures \cite{wolfs2018,wolfs2019}, to more advanced modelling strategies extending the activation approach to element-by-element simulations for complex geometries \cite{ooms2021}, non-planar geometries \cite{nguyen-van2021}, and large-scale structures \cite{jendele2025}. These modelling frameworks have been further advanced through the incorporation of increasingly sophisticated constitutive formulations, accounting for phenomena such as elasto-plasticity with hardening \cite{saif-ur-rehman2026,tripathi2022}, finite-strain viscoelasticity \cite{nedjar2022}, aging and damage evolution \cite{chen2025}, coupled multi-physics interactions \cite{pierre2025}, and environmental effects on material behaviour \cite{gribonval2025}. 

However, despite significant modelling advancements, most solid-based FEM models still rely on the simplifying geometrical assumption that extruded filaments can be represented as idealized rectangular geometries, with dimensions typically derived from the nozzle size or from preliminary measurements of the deposited filament width when available. While the errors introduced by this hypothesis may remain negligible in cases involving extrusion through rectangular nozzles with stiff materials, it has been suggested that, for circular nozzles, this may represent a significant oversimplification, potentially leading to consistently reduced predictive accuracy \cite{liu2021,reinold2024}. 

Accurately representing the actual filament geometry and morphology would therefore require prior knowledge of the deposited shape. Although this information can be obtained through preliminary extrusion tests, such an approach is often impractical and time-consuming, especially under complex printing conditions, and conflicts with the predictive objective of numerical modelling frameworks. A possible alternative, given that the relationships between process parameters, material properties, and resulting filament shape are non-trivial, is to rely on advanced fluid-based numerical models, which have seen significant development in recent years to simulate extrusion and layer deposition processes. In particular, continuum fluid models implemented within the Finite Volume Method (FVM) \cite{comminal2020} and the Particle Finite Element Method (PFEM) \cite{reinold2022, rizzieri2024} have demonstrated high accuracy in predicting filament shapes for single and also few superimposed layers.

These high-fidelity models can resolve the filament scale in detail, but are generally computationally demanding, making them unsuitable for reproducing the entire printing process within a single computational framework. This limitation arises because many available solvers are developed for computational fluid dynamics (CFD) rather than solid mechanics, relying on constitutive models and numerical schemes that are not optimized to efficiently capture the evolving strength of fresh concrete and the structural response of printed elements during deposition.

For this reason, if the objective is to predict the entire 3D printing process and assess buildability, they are typically combined with complementary approaches (e.g., layer-activation FEM) or extended through advanced unified formulations. For example, \cite{reinold2024} employed PFEM to simulate the extrusion of individual layers, which were subsequently superimposed and progressively activated to evaluate the buildability of a rectangular structure. Similarly, \cite{rizzieri2026b} introduced a unified elasto-viscoplastic fluid–solid constitutive model, but with the aim of extending the PFEM framework for simulating 3D printing across multiple process scales, from material extrusion to buildability assessment. In contrast, \cite{an2026} proposed a Smoothed Particle Hydrodynamics (SPH)-based model to capture the extrusion process, combined with a finite element method (FEM) approach incorporating layer activation to evaluate buildability. Nonetheless, these developments remain relatively recent and still involve significant computational costs, together with a high level of expertise required for model implementation and application.

These limitations highlight the need for alternative approaches, beyond experimental measurements and fluid-dynamics-based simulations, to enable rapid estimation of filament shape in 3DCP and potentially enhance the accuracy of standard FEM-based buildability models. For instance, by systematically analysing the influence of printing conditions and material characteristics, design charts have been developed to directly estimate filament width and height for free-flow deposition \cite{rizzieri2025} and layer-pressing strategies \cite{carneau2022} in 3DCP. 

Data-driven methods have also started emerging to establish relationships between process parameters, material properties, and filament morphology. For example, regression-based machine learning models have been employed to predict filament width, height and contact width from printing parameters \cite{alhussain2024}. However, the applicability of the proposed model remains limited to layer-pressing printing conditions and the specific material composition investigated. Based on both material and process parameters, \cite{lao2020} developed a support vector machine learning algorithm trained to predict simplified cross-sectional representations of filament shapes (i.e., flat-topped polygonal profiles) for rectangular-nozzle 3D concrete printing at corners. Recently, \textit{ShapeGen3DCP} \cite{rizzieri2026} was also proposed as a highly general deep learning framework for predicting layer shapes in 3D concrete printing with circular nozzles. The model takes as input both material rheological properties, described through the Bingham's parameters, and key process parameters, and provides an instantaneous prediction of the complete cross-sectional profile of single and double layers, encoded using Fourier descriptors. Since the entire profile geometry is predicted, key geometric characteristics such as filament height, width, and interlayer contact can be directly extracted through simple post-processing operations.

The aim of this paper is to integrate data-driven filament shape prediction approaches into existing FEM-based buildability models for 3DCP, improving their predictive reliability by overcoming the limitations associated with simplified geometric representations. In the following, this integration is referred to as a \textit{geometrically informed} modelling approach. Specifically, a workflow is proposed that couples the filament-shape prediction capabilities of \textit{ShapeGen3DCP} with a layer-activation finite element (FEM) framework for buildability assessment. The framework is used to address three key questions: (i) how filament geometry affects buildability predictions; (ii) how this influence depends on printing strategy, process parameters, and material behaviour; and (iii) what level of geometric fidelity is required to obtain accurate predictions without excessive computational cost.

The article is organized as follows. Section \ref{sec:2} presents the overall workflow and details its three main components: (1) \textit{ShapeGen3DCP}, a deep-learning-based tool for predicting filament geometries from material properties and printing parameters; (2) a mesh-generation procedure that combines the predicted filament shapes with the printing toolpath to obtain a geometrically accurate representation of the printed object; and (3) an elastic and elasto-plastic FEM framework for buildability assessment up to failure, followed by post-processing of the results.
Section \ref{sec:3} presents the results. The workflow is first validated against experimental data on single-layer walls exhibiting elasto-plastic buckling failure, showing excellent agreement when realistic filament geometries are considered. A parametric study is then conducted on rectilinear walls, considering six printing strategies, five geometric discretizations, and three material models. The results quantify the influence of filament shape on buildability and reveal how this influence depends on both process conditions and material behavior. In particular, layer-pressing regimes are found to be less sensitive to geometric simplifications than free-flow deposition conditions. It is also recognized that elliptical cross-sections often provide a good balance between accuracy and computational efficiency, while a volume-conservation-based dimensioning approach is proposed to improve the accuracy of rectangular approximations. The framework is subsequently applied to cylindrical geometries, confirming the trends observed for walls.
Finally, Section \ref{sec:4} summarizes the main findings and presents practical guidelines for incorporating filament geometry into buildability analyses, including a simplified design chart for single-layer 3D-printed walls.

\section{Workflow description}\label{sec:2}
\begin{figure}[b]
    \centering
    \includegraphics[width=1.0\linewidth]{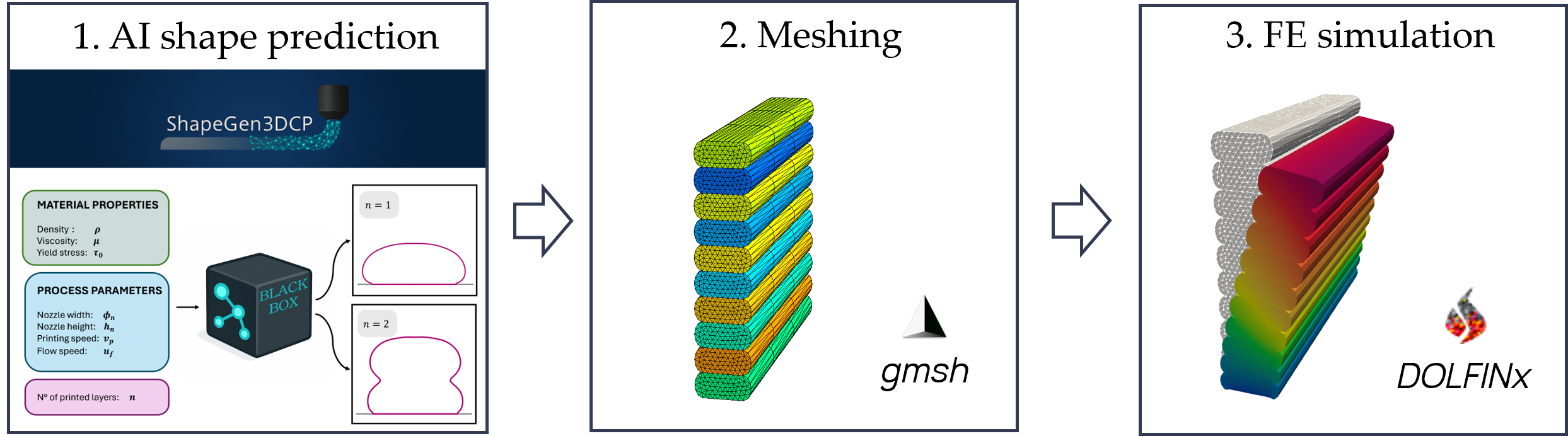}
    \caption{Automated workflow integrating machine-learning based filament shape prediction (\textit{ShapeGen3DCP}) with robust layer-by-layer mesh generation (\textit{gmsh}) and buildability finite element simulations (in \textit{DOLFINx}).}
    \label{fig:workflow}
\end{figure}

This section presents the automated workflow for conducting geometrically informed buildability simulations in extrusion-based 3D printing. The framework is structured into three sequential and interdependent phases, schematically illustrated in Figure \ref{fig:workflow}. Each phase is defined by specified inputs and outputs, enabling consistent data transfer across the pipeline.

\begin{enumerate}
    \item The first phase addresses the generation of filament cross-sectional geometries for single- and double-layer deposition. These geometries represent the outcome of the coupled extrusion and deposition processes. In this work, the cross-sectional profile is predicted using \textit{ShapeGen3DCP}~\cite{rizzieri2026}, a recently developed deep learning framework capable of rapidly estimating filament shapes from process parameters (e.g., extrusion rate, nozzle speed) and material properties in the fluid state. This approach enables efficient and physically informed reconstruction of deposited filament geometries without resorting to computationally expensive multiphysics simulations.

    \item The predicted cross-sectional information is then passed to the second phase, where the goal is to construct a mesh representation of the overall printed object. At this stage, it is not necessary to strictly preserve the exact geometry predicted by \textit{ShapeGen3DCP}. Instead, the focus shifts to creating simplified but mechanically meaningful approximations of the filament cross-section that can be used as building blocks for the layered computational mesh. For instance, the \textit{ShapeGen3DCP} outputs can be used to define idealized shapes such as rectangular or elliptical profiles. This representative cross-section is then swept along the prescribed toolpath and replicated vertically to form a three-dimensional model of the printed object. Finally, the model is discretized into a tetrahedral finite element mesh using \textit{Gmsh}, making it suitable for subsequent numerical analysis.

    \item In the third phase, the generated computational domain is imported into a simulation environment for mechanical assessment. Buildability is evaluated through a finite element analysis framework in which layers are progressively activated to mimic the sequential nature of the printing process. A quasi-static elastoplastic formulation is employed to capture the accumulation of deformation and the onset of failure as the structure evolves during fabrication.
\end{enumerate}

The pipeline is implemented in Python and managed using the rule-based workflow management system Snakemake \cite{molder2021}, ensuring reproducible and automated execution of all processing steps. Each phase is executed in an isolated environment tailored to its dependencies -- for instance, PyTorch for machine-learning-based filament cross-section prediction, \textit{Gmsh}~\cite{geuzaine2009} for geometry and mesh generation, and the DOLFINx~\cite{baratta2023} framework for finite element analysis. In this way, completed tasks are tracked automatically, and failed or outdated steps can be selectively recomputed without rerunning the entire pipeline. Still, the modular architecture allows individual workflow stages to be executed independently, facilitating debugging and the verification of intermediate results. The implementation is available on Zenodo \cite{rizzieri2026a}.

\subsection{\textit{ShapeGen3DCP}: machine-learning predictions of filament shape outline}\label{sec:shapegen}
Accurate prediction of single- and multi-layer filament cross-sectional shapes has been extensively investigated in the literature using fluid-based finite volume \cite{comminal2020} and finite element \cite{rizzieri2024,rizzieri2025} approaches. While these techniques yield high-fidelity results, they are computationally intensive -- even for a few layers -- and require substantial setup, limiting their use in integrated or real-time workflows.

To address these limitations, the present study leverages \textit{ShapeGen3DCP} \cite{rizzieri2026}, a machine learning framework recently introduced in the literature for the prediction of filament cross-sectional shapes. The model takes as input both material properties in the fluid state -- namely density ($\rho$), yield stress ($\tau_0$), and plastic viscosity ($\mu$) -- and key process parameters, including nozzle diameter ($\phi_n$), nozzle height ($h_n$), printing velocity ($v_P$), and flow velocity ($u_f$).

\begin{figure}[t]
    \centering    \includegraphics[width=0.9\linewidth]{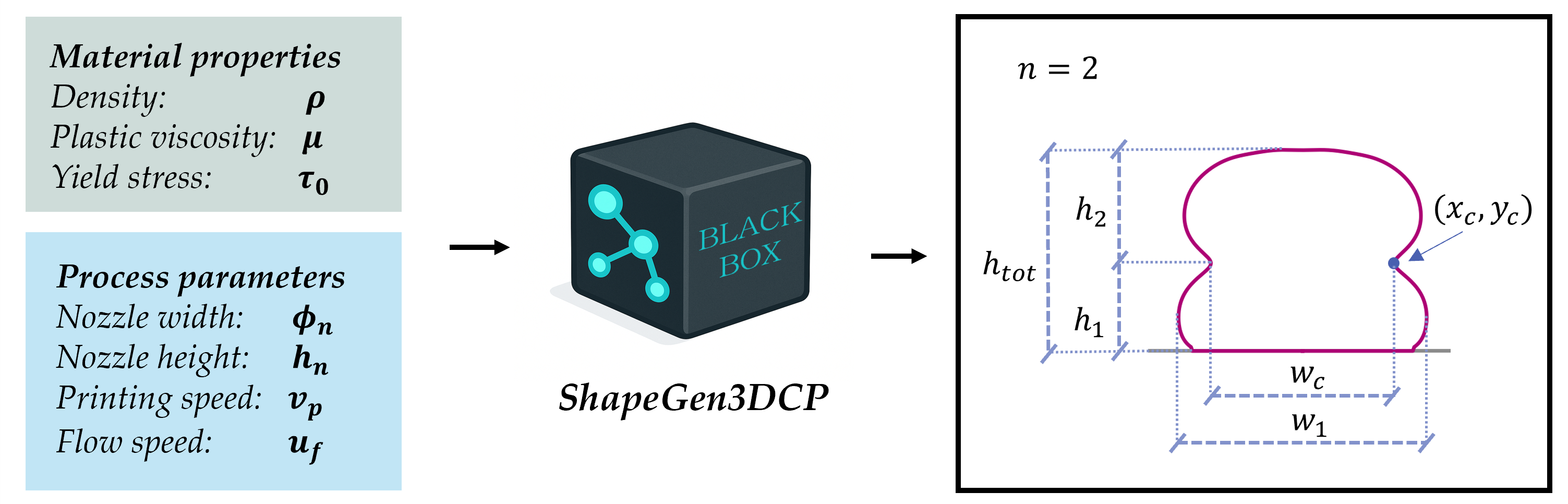}
    \caption{\textit{ShapeGen3DCP} overview for the two-layer case: material properties in the fluid state and process parameters are provided as inputs, yielding a smooth, closed, symmetric cross-sectional profile. The drawing further highlights the key geometrical features that can be extracted from the resulting shape.}
    \label{fig:shapegen3dcp}
\end{figure}

\textit{ShapeGen3DCP} was trained on a numerically generated dataset obtained using a validated Particle Finite Element Method (PFEM) for simulating additive manufacturing with cementitious materials \cite{rizzieri2024}. The dataset includes filament geometries for approximately 180 different combinations of material and process parameters. A deep neural network maps these input parameters to a parametric representation of the filament cross-sections using Fourier descriptors, which inherently preserve geometric properties such as smoothness, closure, and symmetry (Figure~\ref{fig:shapegen3dcp}). Importantly, validation against experimental data from multiple independent sources and the full-order PFEM model showed shape-prediction errors generally within 1–10\% for key geometric descriptors (e.g., maximum width, interlayer contact, height, cross-sectional area) with only a few cases reaching approximately 15\%, demonstrating the accuracy and robustness of \textit{ShapeGen3DCP} for advanced applications. For a detailed description of \textit{ShapeGen3DCP} architecture, training procedure and validation the reader is referred to \cite{rizzieri2026b}.

The model can predict both single-layer and two-layer extrusion configurations. In this study, the focus is on the latter, as two-layer cross-sections provide a detailed representation of interlayer interaction and merging behaviour. In addition, \textit{ShapeGen3DCP} enables the post-processing of reconstructed filament shapes to extract reduced geometric descriptors, as illustrated in Figure~\ref{fig:shapegen3dcp}.

For two-layer cross-sections, the relevant descriptors include the bottom layer's maximum width $w_1$, the interlayer contact length $w_c$, the bottom layer height $h_1$, the top layer height $h_2$, the total height $h_{tot}=h_1+h_{tot}$, and the total cross-sectional area $A$, computed numerically using the shoelace formula applied to the reconstructed contour.

The interlayer contact length $w_c$ and the bottom layer height $h_1$ are determined by locating the \emph{interlayer contact point} $(x_c, y_c)$, defined as the point along the contour where the first derivative -- approximated via finite differences -- changes sign. The horizontal coordinate of this point gives half of the contact length, so that $w_c = 2x_c$, while the vertical coordinate directly provides the bottom layer height, i.e.\ $h_1 = y_c$.

\subsection{Filament shape post-processing and meshing strategy}\label{sec:filament}
Numerical models for 3DCP buildability assessment often rely on simplified assumptions regarding layer geometry. This simplification is typically driven by limited prior knowledge of the actual filament cross-section, as well as the need to reduce mesh complexity and computational cost. However, as demonstrated below, filament shape can significantly influence buildability predictions. It is therefore essential to adopt a representative cross-section that achieves an appropriate balance between geometrical fidelity and model complexity. Furthermore, even in the absence of experimental measurements, filament geometry can now be efficiently estimated using data-driven approaches, such as the method introduced in the previous section~\ref{sec:shapegen}.

\subsubsection{Representative layer shape}\label{sec:layershapes}
\begin{figure}[h]
    \centering
    \includegraphics[width=1.0\linewidth]{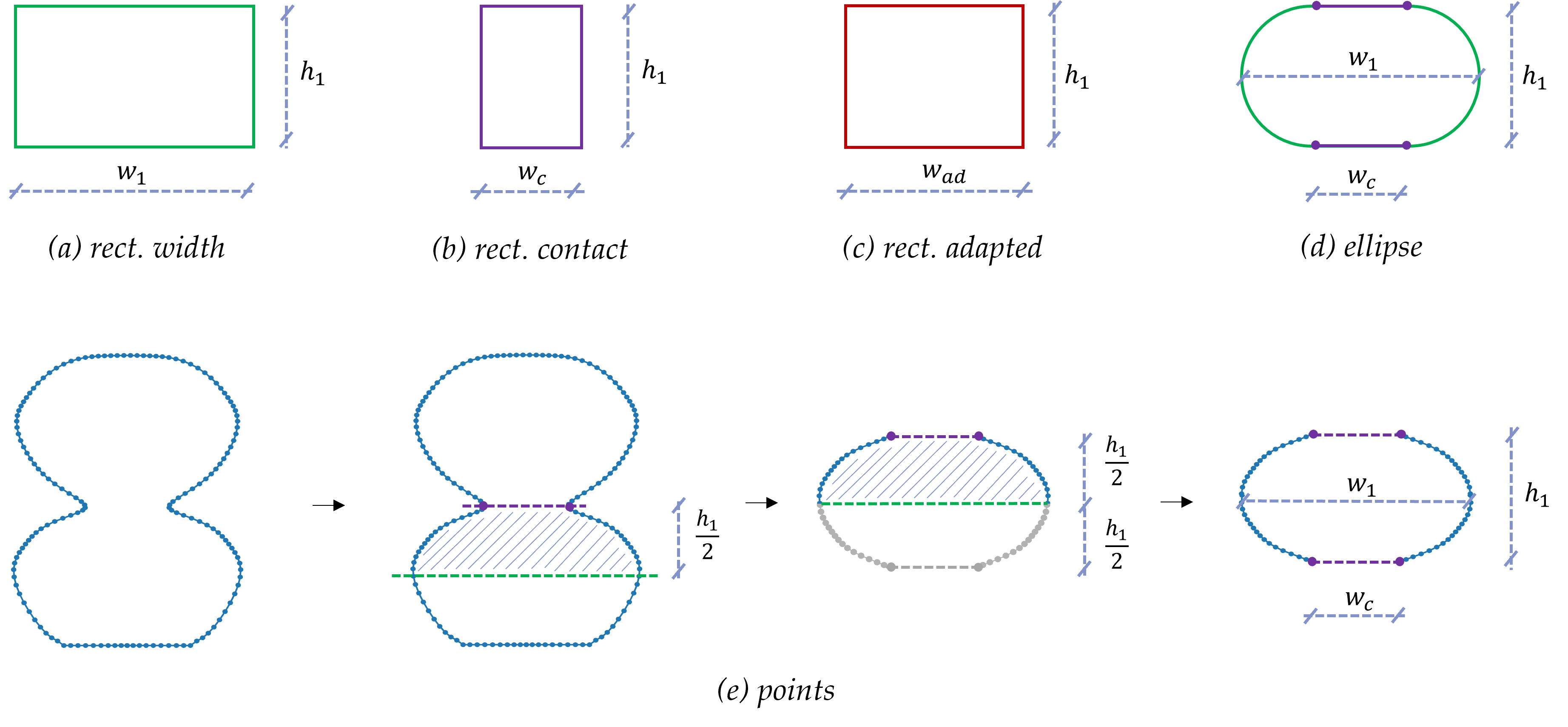}
    \caption{Different representative layer geometries with increasing geometric fidelity: (a) \textit{rect. width}, (b) \textit{rect. contact}, (c) \textit{rect. adapted}, (d) \textit{ellipse}, and (e) \textit{points}.}
    \label{fig:layer-shapes}
\end{figure}

In this context, the present study adopts  --  as a reference  -- the two-layer cross-sectional geometry predicted by \textit{ShapeGen3DCP}. The bottom layer is of particular interest, as it represents a filament that has not only been deposited but has also undergone deformation due to the localized pressure increase caused by material inflow during the extrusion of the subsequent layer. Starting from this configuration, different representative layer geometries are derived, corresponding to increasing levels of geometric fidelity, as illustrated in Figure~\ref{fig:layer-shapes}.

\begin{itemize}
    \item \emph{Full-width rectangle (\textit{rect. width}).} 
    The cross-section is idealized as a rectangle of width equal to the bottom layer's maximum width ($w = w_1$) and height equal to the bottom-layer height ($h = h_1$); this simplification generally leads to an overestimation of the actual extruded material volume.

    \item \emph{Contact-width rectangle (\textit{rect. contact}).} The cross-section is idealized as a rectangle of width equal to the interlayer contact length ($w = w_c$) and height equal to the bottom-layer height ($h = h_1$); this assumption generally underestimates the actual extruded material volume.

    \item \emph{Area-equivalent rectangle (\textit{rect. adapted}).} 
    The cross-section is idealized as a rectangle with prescribed height ($h = h_1$), while the width is adjusted to preserve the cross-sectional area of the point-derived profile (covered in the following), i.e., $w_{\mathrm{ad}} = A_{points} / h_1$. This construction ensures exact conservation of the extruded material volume.

    \item \emph{Elliptical-capped rectangle (\textit{ellipse}).} 
    The cross-section is modelled as a composite geometry consisting of a central rectangular core of width $w_c$ and height $h_1$, flanked by two lateral regions defined by elliptical segments. The overall width of the cross-section is $w_1$.

    \item \emph{Point-derived profile (\textit{points}).} 
    The cross-section is constructed directly from the \textit{ShapeGen3DCP} point-wise profile prediction. The bottom layer is considered as it already encodes deformation effects induced by subsequent layer extrusion and deposition. To remove the influence of the rigid substrate, only the upper portion of the profile is retained (points with $y > 0.5h_1$). This partial profile is then mirrored about the $y$-axis to obtain a symmetric cross-section suitable for vertical stacking. The final geometry is a closed polyline with a user-defined number of points, bounded by $w_1$ in width and $h_1$ in height, with horizontal top and bottom segments of length $w_c$ and an enclosed area of $A_{points}$.
\end{itemize}

\subsubsection{Mesh generation}\label{sec:meshing}
The geometrical descriptions introduced in the previous section are converted into finite element meshes through an automated pipeline based on \textit{Gmsh}. The adoption of \textit{Gmsh} is motivated by its capability to construct geometries programmatically, handle complex curve definitions, and generate consistent three-dimensional unstructured meshes from parametrized inputs. The meshing procedure follows a sequential and modular workflow, which is consistent across all filament representations.

\begin{figure}[t]
    \centering
    \subfloat[\textit{rect. width}]{\includegraphics[width=0.14\linewidth]{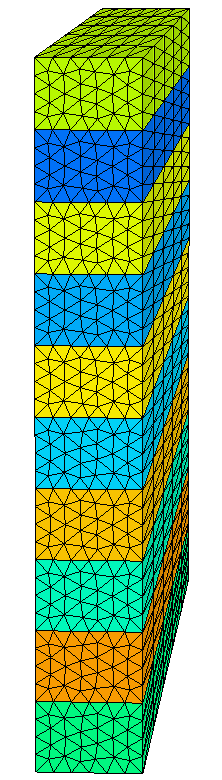}}\hspace{0.8cm}
    \subfloat[\textit{rect. adapted}]{\includegraphics[width=0.14\linewidth]{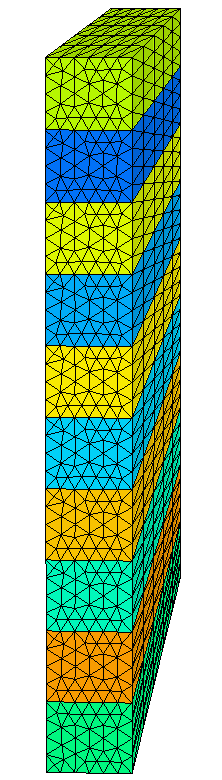}}\hspace{0.8cm}
    \subfloat[\textit{rect. contact}]{\includegraphics[width=0.14\linewidth]{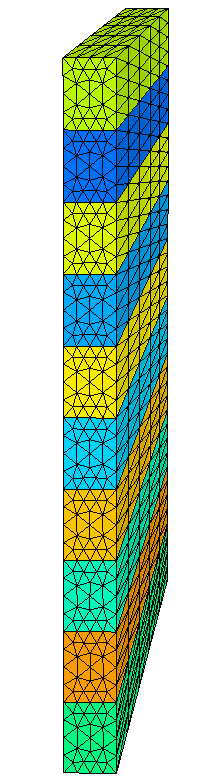}}\hspace{0.8cm}
    \subfloat[\textit{ellipse}]{\includegraphics[width=0.14\linewidth]{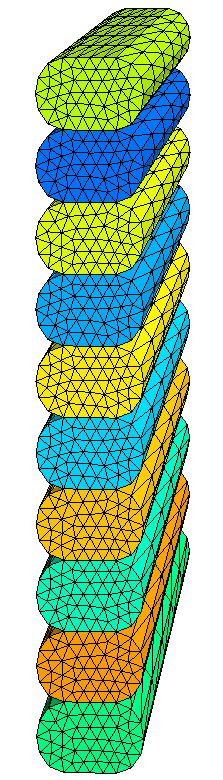}}\hspace{0.8cm}
    \subfloat[\textit{points}]{\includegraphics[width=0.14\linewidth]{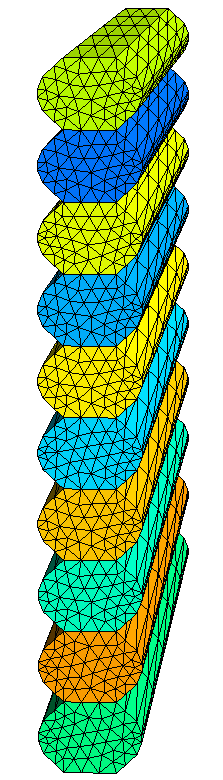}}
    \caption{Meshed computational domains generated in \textit{Gmsh} by extruding reference layer geometries at varying accuracy levels and subsequently stacking the resulting layers vertically.}
    \label{fig:gmsh}
\end{figure}

\begin{enumerate}
    \item \emph{Definition of the cross-sectional outline.} The starting point is the definition of a closed two-dimensional profile representing the filament cross-section. Depending on the chosen representation, this profile is either generated analytically (rectangular and elliptical cases) or directly derived from processed point data (point-based case).

    For the rectangular and elliptical geometries, a set of characteristic points is first defined from the prescribed geometric parameters (layer height, layer width, and contact width). These points are then connected to form the cross-section: by straight segments for the rectangular variants (\textit{rect. width}, \textit{rect. contact}, \textit{rect. adapted}), and by a combination of straight segments and elliptical arcs along the lateral boundaries for the elliptical case (\textit{ellipse}), allowing a smoother representation of the filament curvature.

    For the point-based representation (\textit{points}), the profile is directly derived from the \textit{ShapeGen3DCP} output after preprocessing, which includes the removal of duplicate points and the enforcement of symmetry. The resulting points are then connected by straight line segments to form the cross-sectional profile.

    In all cases, the resulting set of lines and curves defining the closed profile is assembled into a curve loop, which is subsequently used to define a planar surface representing the single-layer cross-section.

    \item \emph{Generation of a single three-dimensional filament.}
    Once the two-dimensional surface is defined, it is transformed into the filament three-dimensional volume. Two alternative strategies are considered, depending on the target geometry.

    For prismatic wall-like structures, the surface is extruded along the printing direction. The extrusion length corresponds to the filament length, and the discretization along this direction can be explicitly prescribed.
    
    For axisymmetric configurations (e.g., cylindrical geometries), the same two-dimensional cross-section is instead revolved around a prescribed axis. The revolution is performed over a full angular span, generating a closed three-dimensional layer. Here, the same \textit{ShapeGen3DCP} cross-section based on straight filaments is used directly, neglecting potential effects of the curvature on the filament geometry and volume.

    \item  \emph{Layer-by-layer volume domain generation.}
    The extrusion (or revolution) operations described above can be applied iteratively, with each step beginning from a surface profile that has been translated along the z‑axis by the inter‑layer distance. This process builds the final stacked geometry layer by layer, ensuring consistent vertical alignment and precise definition of the three‑dimensional domain.

    \item \emph{Mesh generation and layer-by-layer activation.}
    After completion of the geometric construction, a three-dimensional unstructured mesh is generated within \textit{Gmsh}. In this work, only tetrahedral elements are used, providing robustness for arbitrary geometries without imposing constraints associated with structured discretizations.
    Figure~\ref{fig:gmsh} shows representative meshes corresponding to the different cross-sectional profiles defined in Section~\ref{sec:layershapes}.

    A key aspect of this approach is the preservation of geometrical interfaces between consecutive layers. As each layer is created as an independent volume, interlayer boundaries are naturally embedded in the mesh. This is particularly beneficial for finite element simulations, as it allows a direct and consistent implementation of layer-by-layer activation without additional geometric manipulation.

\end{enumerate}

\subsection{Simulation framework}\label{sec:simulation}
The simulation framework adopts the general assumption, first proposed in \cite{wolfs2018}, that freshly deposited 3D printable concrete can be modelled as a continuum solid. In this work, owing to the proposed integrated workflow, the effects of material flow during extrusion have already been accounted for in the computational mesh through the shape of the filaments. The simulation framework is implemented using the open-source finite element software DOLFINx~\cite{baratta2023} and uses the framework presented in~\cite{rosenbusch2025} to integrate the user-defined constitutive model.

\subsubsection{Governing equations and numerical formulation}
The structural response during printing is governed by the quasi-static balance of linear momentum,
\begin{equation}
    \nabla \cdot \big[\chi(\boldsymbol{x},t)\,\boldsymbol{\sigma(x,t)}\big] + \chi(\boldsymbol{x},t)\,\rho\,\boldsymbol{g} = \boldsymbol{0},
    \label{eq:equilibrium}
\end{equation}
where $\boldsymbol{\sigma}$ is the Cauchy stress, $\rho$ the constant material density and $\boldsymbol{g}$ the gravity vector. 
The full mesh of the target geometry is built a priori (Section~\ref{sec:meshing}), and the progressive layer-by-layer deposition is reproduced through a time-dependent pseudo-density field $\chi(\boldsymbol{x},t)\in[0,1]$ defined over the entire domain, which scales both the stress contribution -- and thereby the stiffness, since the consistent tangent is $\chi\,\partial\boldsymbol{\sigma}/\partial\boldsymbol{\varepsilon}$ -- and the body force in eq.~\eqref{eq:equilibrium}. For $\chi=0$, the material contributes neither stiffness nor weight; for $\chi=1$ it responds as a fully active solid. In the sub-domain corresponding to the $i$-th deposited layer, $\chi$ rises continuously from 0 to 1 over the associated layer-activation interval. 

Fresh, early-age concrete often exhibits a comparatively low internal friction angle (on the order of a few degrees) after deposition \cite{suiker2020, wolfs2019a}, leading to negligible difference between pressure-sensitive (e.g. Mohr-Coulomb) and pressure-independent yield criteria in this regime \cite{suiker2020}.
Furthermore, fluid-based simulations of the deposition process standardly rely on pressure-independent viscoplastic (Bingham-type, von Mises) formulations \cite{reinold2024, rizzieri2024}. For these reasons and since the focus of the work is the layer shape influence, the deposited material is described by a $J_2$ elasto-plastic constitutive law with non-linear isotropic (saturation-type) hardening, following the formulation introduced in \cite{saif-ur-rehman2026}.
With the additive decomposition $\boldsymbol{\varepsilon}=\boldsymbol{\varepsilon}^e+\boldsymbol{\varepsilon}^p$ of the strain into elastic and plastic parts,  the model is derived from the Helmholtz free energy per unit volume, additively split into an elastic and a stored plastic (hardening) contribution,
\begin{equation}
\psi(\boldsymbol{\varepsilon}^e,\alpha)
= \hat{\psi}^e(\boldsymbol{\varepsilon}^e) + \hat{\psi}^p(\alpha)
= \tfrac{1}{2}\,\kappa\,\varepsilon_v^2
+ \mu\,\boldsymbol{\varepsilon}^{e\prime}\!:\!\boldsymbol{\varepsilon}^{e\prime}
+ (\tau_\infty - \tau_0)\left( -\frac{1}{\omega} + \alpha
+ \frac{1}{\omega}e^{-\omega\,\alpha} \right),
\label{eq:free_energy}
\end{equation}
where $\kappa$ and $\mu$ are the bulk and shear moduli, $\varepsilon_v=\mathrm{tr}\,\boldsymbol{\varepsilon}^e$ the volumetric elastic strain and $\boldsymbol{\varepsilon}^{e\prime}$ its deviatoric part, $\alpha$ the accumulated plastic strain, $\tau_0$ and $\tau_\infty$ the initial and saturation yield limits, and $\omega$ the hardening factor controlling the transition between $\tau_0$ and $\tau_\infty$. Standard Coleman--Noll arguments give the stress $\boldsymbol{\sigma}=\partial\psi/\partial\boldsymbol{\varepsilon}^e=\kappa\,\varepsilon_v\,\boldsymbol{1}+2\mu\,\boldsymbol{\varepsilon}^{e\prime}$ and the hardening stress conjugate to $\alpha$, $q=\partial\hat{\psi}^p/\partial\alpha=(\tau_\infty-\tau_0)\,(1 - e^{-\omega\,\alpha})$. The yield function then reads,
\begin{equation}
    \phi(\boldsymbol{\sigma},\alpha) = \|\boldsymbol{\sigma}'\| - \sqrt{\tfrac{2}{3}}\,
    \big[\,\tau_0 + (\tau_\infty - \tau_0)\,(1 - e^{-\omega\,\alpha})\,\big],
    \label{eq:yield_vM}
\end{equation}
where $\boldsymbol{\sigma}'$ is the deviatoric Cauchy stress. The flow rule is associative, giving $\dot{\boldsymbol{\varepsilon}}^p = \lambda\,\boldsymbol{n}$ and $\dot{\alpha} = \sqrt{2/3}\,\lambda$, with the unit deviatoric flow direction $\boldsymbol{n}=\boldsymbol{\sigma}'/\|\boldsymbol{\sigma}'\|$ and the plastic multiplier $\lambda\geq 0$ subjects to the Karush--Kuhn--Tucker conditions $\lambda\geq 0$, $\phi\leq 0$, $\lambda\,\phi=0$. 

For the numerical solution, the quasi-static problem eq.~\eqref{eq:equilibrium} is transformed into its weak form and discretized in space using the finite element method with appropriate trial and test functions. Owing to the nonlinear constitutive response, the resulting system of equations is solved iteratively at each load step of the quasi-static analysis using a standard Newton--Raphson solver.
At each quadrature point and each load step, the constitutive update is integrated implicitly (backward Euler) by a radial-return scheme: starting from the elastic trial stress, the yield criterion is checked and, if violated, a local Newton iteration on $\lambda$ enforces consistency ($\phi=0$); the stress, plastic strain and hardening variable are then updated and the consistent algorithmic tangent is returned to the global Newton--Raphson solver to preserve quadratic convergence.

Because the layer-by-layer build-up 
produces moderate-to-large rotations of the deposited material, the simulations are carried out within an updated Lagrangian framework with objective stress rates \cite{hughes1980,saif-ur-rehman2026arxiv}. 
The reference configuration is advanced to the current deformed configuration at the end of every converged load step, so that all subsequent quantities are referred to the latest mesh. Within each increment, the Cauchy stress is transported objectively through the Jaumann (co-rotational) rate: the skew-symmetric part of the displacement-gradient increment, $\Delta\boldsymbol{W} = \mathrm{skw}(\nabla\Delta\boldsymbol{u})$, defines a Hughes-Winget incremental rotation tensor, which is used to push the previous stress state forward before the constitutive update is evaluated at the midpoint configuration. This procedure prevents spurious stress accumulation under rigid-body rotations and keeps the material state consistent with the structural deformation.  

Finally, the structural build-up and ageing of fresh concrete during printing are taken into account by letting each material parameter $P\in\{E, \nu, \tau_0, \tau_\infty, \omega\}$ (with the elasticity parameters: Young's modulus $E$, Poisson ratio $\nu$) vary linearly with the simulation time,
\begin{equation}
    P(t) = P_0 + A_P\,t,
    \label{eq:time_dep_params}
\end{equation}
with initial value $P_0$ at $t=0$ and evolution rate $A_P$. Both quantities are chosen from compression tests performed on samples of different ages, as discussed in Section~\ref{sec:validation}, providing the gradual gain in stiffness and strength necessary to reproduce the experimentally observed buildability. The linear evolution is a common assumption in buildability simulations  \cite{suiker2018, wolfs2018, ooms2021, reinold2024, nedjar2022}. Owing to the relatively short time scale relevant for the investigated buildability analysis, in the order of minutes, the material remains in the dormant phase, such that a first-order linear approximation is  sufficient~\cite{roussel2012}.
%

\subsubsection{Simulation set-up for buildability analysis}
\begin{figure}[h]
    \centering
    \includegraphics[width=0.2\linewidth]{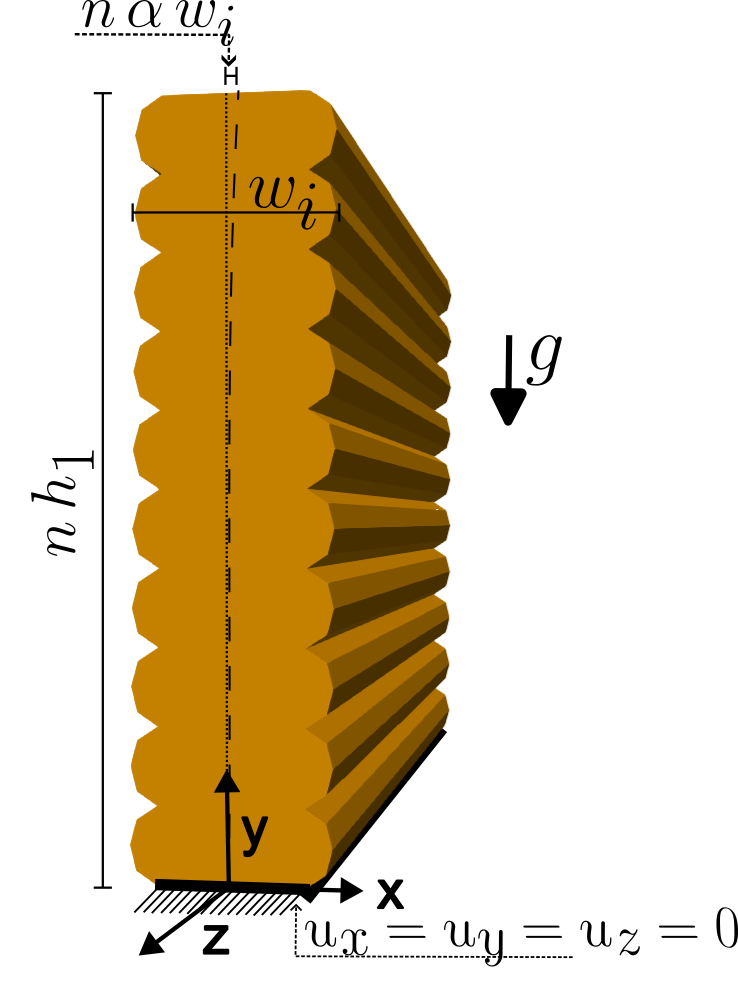}
    \caption{Geometry and boundary conditions for buildability simulation at the example of printing a wall.}
    \label{fig:wall}
\end{figure}

For the buildability simulations considered in the following, the layer cross-section is defined in the $x$--$y$ plane and extruded in the $z$-direction, as illustrated for the wall example in Figure~\ref{fig:wall}. The base of the first layer is fixed in all spatial directions. Gravity ($g=9.81$~m/s$^2$) is applied in the negative $y$-direction. The load contribution of each newly deposited layer is increased linearly over the activation time of one layer, $t_l = l_l/v_p$,
where $l_l$ denotes the layer length and $v_p$ the printing velocity. 
 This linear ramp accounts for the fact that, in reality, material is deposited continuously over the extrusion length, while in the simulation, all elements of a layer are activated simultaneously. Thus, the increase in self-weight during deposition is approximated within the layer activation interval. The load step assumes a dual role in the present formulation. It constitutes the physical increment over which the age-dependent stiffness and the deposition of self-weight are advanced. Through the load imposed within that interval it further prescribes the strain and rotation increments upon which the accuracy of the rate-independent return mapping and of the objective stress update depends.

To trigger the first buckling mode, a linear geometric imperfection is introduced over the height of the structure in the out-of-plane direction ($x$). The imperfection amplitude is defined per layer as a multiple $\alpha$ of the layer width $w_i$ ($w_1$, $w_c$, or $w_{ad}$, depending on the adopted filament representation defined in Sect.~\ref{sec:filament}).

Failure of the printed structures is identified using a displacement-based criterion. The simulation is terminated when any active node reaches an out-of-plane displacement exceeding a prescribed threshold. In the following, a threshold of $0.1$ times the layer width is used for the wall simulations based on a comparison with the analytical solution in the elastic case. For cylinder simulations, a threshold of one layer width is used as in~\cite{saif-ur-rehman2026}. Other stopping criteria based on eigenvalue analysis are discussed in \cite{saif-ur-rehman2026arxiv}.
The corresponding number of activated layers at failure is then computed from the ratio of the failure time $t_{\mathrm{fail}}$ to the layer activation time:
\begin{equation}\label{eq:num_activated_layers}
    n_{\mathrm{act}} = \frac{t_{\mathrm{fail}}}{t_l}.
\end{equation}

Finally, it is worth mentioning that different filament representations result in different total structural volumes, which in turn affect the applied load and, consequently, the simulation results. Only the \textit{points} and \textit{rect. adapted} representations are volume-preserving, whereas the others aim to preserve the geometrical characteristics predicted from \textit{ShapeGen3DCP}. The material density can be adjusted so that all structures have the same total weight. The effect of such a density adjustment on the buildability is discussed in the validation example in Sect.~\ref{sec:validation}.

\section{Results}\label{sec:3}
This section first evaluates the proposed workflow by examining its predictions against published experimental observations from the literature, assessing its ability to predict buildability. The framework is then employed to investigate the influence of filament shape and its geometrical discretization on the buildability of rectilinear walls under different material behaviours, ranging from purely elastic to various elasto-plastic regimes. Finally, buildability is assessed for cylindrical structures to identify the combined effects of filament shape and global structural geometry.

\subsection{Validation of geometry-informed approach}\label{sec:validation}
As an initial assessment of the methodology proposed in Section~\ref{sec:2}, the simulation results are examined against printing experiments reported by Tripathi et al.~\cite{tripathi2022}. This study was selected because it provides all the data required for the present analysis, including the printing parameters and the fluid and solid material properties. Moreover, it considers multiple material compositions and reports extensive buildability experiments on simple wall geometries, providing a suitable basis for assessing the predictions of the proposed framework for two distinct materials.

\begin{figure}[bt]
    \centering
    \subfloat[Material LM 30]{\includegraphics[width=0.45\linewidth]{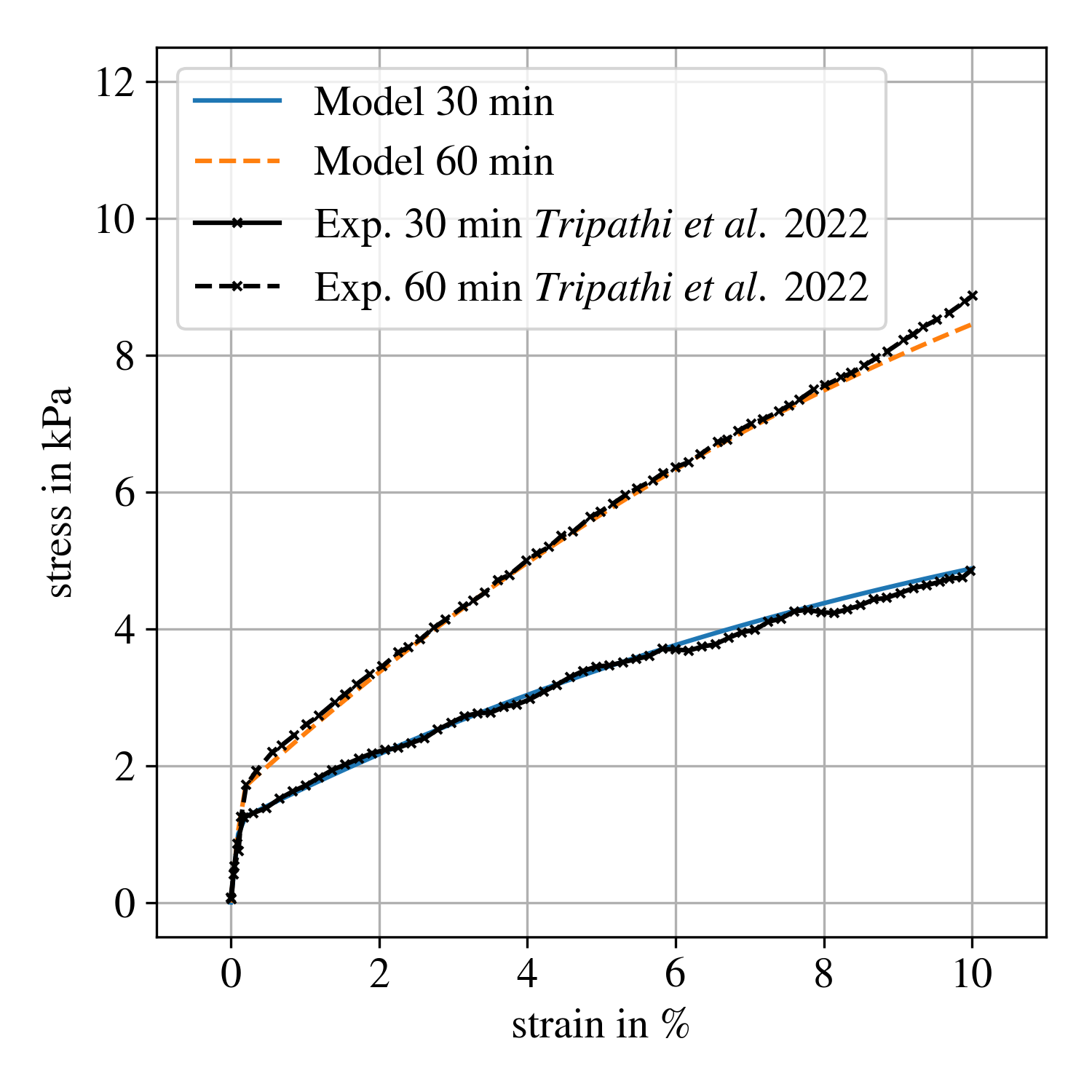}} %
    \subfloat[Material LSM 30]{\includegraphics[width=0.45\linewidth]{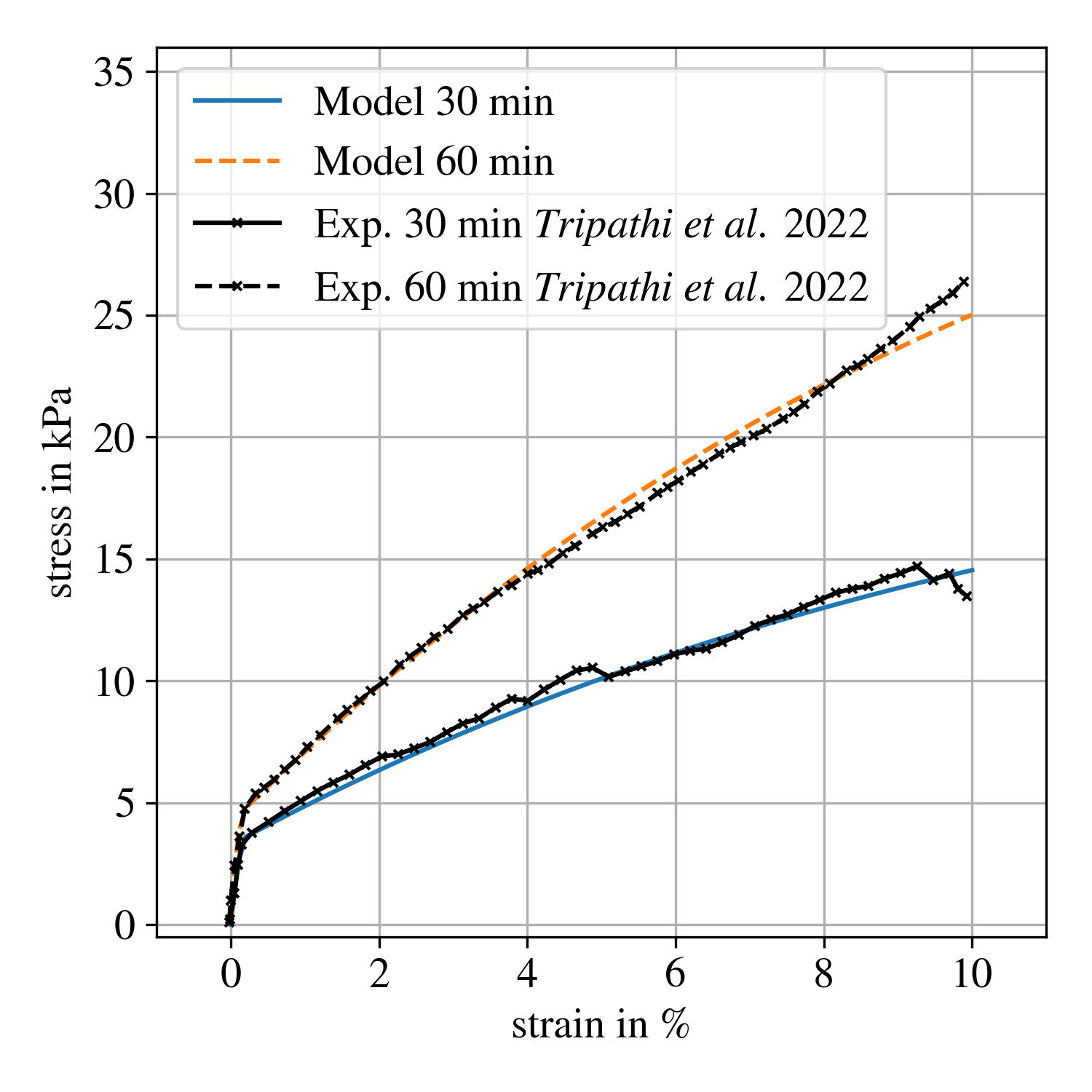}}
    \caption{Stress-strain curves from a uniaxial cube simulation under compression compared with the experimentally measured stress-strain curves from \cite{tripathi2022}.}
    \label{fig:calibration}
\end{figure}

First, the parameters of the elastoplastic material model with nonlinear hardening are chosen using the stress-strain curves obtained from compression tests on samples of different ages reported by Tripathi et al. For simplicity, these compression tests are reproduced numerically as uniaxial, displacement-controlled compression simulations. 
The material parameters are selected to match the reported stress-strain curves for the two materials LM~30 and LSM~30 from Tripathi et al. for two ages, assuming the common linear time evaluation in between (see eq.~\eqref{eq:time_dep_params}). Figure~\ref{fig:calibration} compares the experimental and simulated stress-strain responses and shows close agreement. The material parameters used in the subsequent analyses are summarized in Table~\ref{tab:material_parameters}.

\begin{table}[h]
\caption{Material parameters and evolution rates for the elastoplastic material law with nonlinear hardening (Young's modulus $E$, Poisson ratio $\nu$, initial yield stress $\tau_0$, final yield stress $\tau_{\infty}$, hardening factor $\omega$ and their corresponding evolution rates $A_{p}$).}
\label{tab:material_parameters}
\centering
\begin{tabular}{lcccccccccc}
\toprule
\textbf{material} &
\makecell{$E$ \\ (Pa)} &
\makecell{$A_{E}$ \\ (Pa/s)} &
\makecell{$\nu$ \\ (-)} &
\makecell{$A_{\nu}$ \\ (1/s)} &
\makecell{$\tau_0$ \\ (Pa)} &
\makecell{$A_{\tau_0}$ \\ (Pa/s)} &
\makecell{$\tau_\infty$ \\ (Pa)} &
\makecell{$A_{\tau_\infty}$ \\ (Pa/s)} &
\makecell{$\omega$ \\ (-)} &
\makecell{$A_\omega$ \\ (1/s)} \\
\midrule
LM 30  & 989999.4  & 22.11  & 0.3  & 0.  & 1210.  & 0.25  & 7000.  & 3.1  & 10. & 0. \\
LSM 30 & 2509998.4 & 650.  & 0.3  & 0. & 3490.  & 0.583 & 21000. & 9.   & 10. & 0. \\
\bottomrule
\end{tabular}
\end{table}

Second, the complete simulation workflow is executed. The filament cross-sectional geometry is predicted using \textit{ShapeGen3DCP} for a nozzle diameter of $20$~mm, a nozzle height of $15$~mm, a printing velocity of $v_p = 15$~mm/s, a flow velocity of $u_f = 17.92$~mm/s, and a density of $2112.5$~kg/m$^3$, as reported by Tripathi et al. The shear yield stress is derived from the initial yield stress by dividing it by $\sqrt{3}$ according to von Mises assumption, and the viscosity is assumed to be $15$~Pa$\cdot$s. For LM~30, the resulting cross-section yields a layer width of $30.3$~mm, an interlayer contact length of $22.13$~mm, and a layer height of $11.7$~mm. For LSM~30, the corresponding values are $27.63$~mm, $19.22$~mm, and $12.21$~mm, respectively. These dimensions are consistent with the approximate generic layer size of $25$~mm $\times$ $15$~mm reported by Tripathi et al., although no material-specific distinction is provided there.

Subsequently, meshes of the $200$~mm long wall specimens are generated using approximately four linear tetrahedral finite elements over one layer height for each of the filament representations introduced in Section~\ref{sec:filament}. For each approach, the printing simulation described in Section~\ref{sec:simulation} is carried out with fixed displacements at the base, an imperfection magnitude of $\alpha = 0.00025$, and a load step of $0.25$~s, which was determined through a convergence study. 

As discussed in Sect.~\ref{sec:simulation}, the different filament representations result in different total volumes. Only the \textit{points} and \textit{rect. adapted} representations are volume-preserving, whereas the remaining approaches preserve specific geometric characteristics predicted by \textit{ShapeGen3DCP}. To assess the influence of this difference, two simulation set-ups are considered. In the first set-up, the material density is adjusted such that all structures have the same total weight and therefore represent the same amount of deposited material. The \textit{points} representation is used as the reference case, and the density is scaled by the factor $A_{\mathrm{points}}/A_i$, with $i \in \{$\textit{rect. width}, \textit{rect. contact}, \textit{rect. adapted}, \textit{ellipse}, \textit{points}$\}$. In the second set-up, no density adjustment is applied. Consequently, the different filament representations lead to different total structural masses, reflecting the volume differences inherent to the respective geometric approximations.

\subsubsection{Validation with experimental data}
Table~\ref{tab:actlayers_wall_tripathi} summarizes the number of activated layers at failure, as defined by Eq.~\eqref{eq:num_activated_layers}, for both set-ups (with and without density adjustment) and compares these values with the experimentally observed numbers of printable layers reported by Tripathi et al. Overall, good agreement is obtained between simulation and experiment, which supports the validity of the proposed framework.

\begin{table}[h]
\caption{Comparison of activated layers at failure for the two materials and varying filament representations. The simulations are run with (w/) and without (w/o) density adjustment. For validation, the number of layers at failure reported in \cite{tripathi2022} is given.}
\label{tab:actlayers_wall_tripathi}
\begin{tabular}{lrrrrrrr}
\toprule
 & & \textit{rect. width} & \textit{rect. adapted} & \textit{rect. contact} & \textit{ellipse} & \textit{points} & exp. from \cite{tripathi2022} \\
material & density adjustment &  &  &  &  &  &  \\
\midrule
LM 30 & w/ & 13.69 & 11.92 & 9.90 & 11.60 & 11.65 & 8 $\pm$ 1 \\
LM 30 & w/o & 13.04 & 11.92 & 10.71 & 11.40 & 11.67 & 8 $\pm$ 1 \\
\midrule
LSM 30 & w/ & 22.08 & 19.27 & 15.50 & 18.56 & 18.60 & 16 $\pm$ 2 \\
LSM 30 & w/o & 20.10 & 19.27 & 18.23 & 17.90 & 18.62 & 16 $\pm$ 2 \\

\bottomrule

\end{tabular}
\end{table}

\begin{figure}[b]
    \centering
    \includegraphics[width=0.95\linewidth]{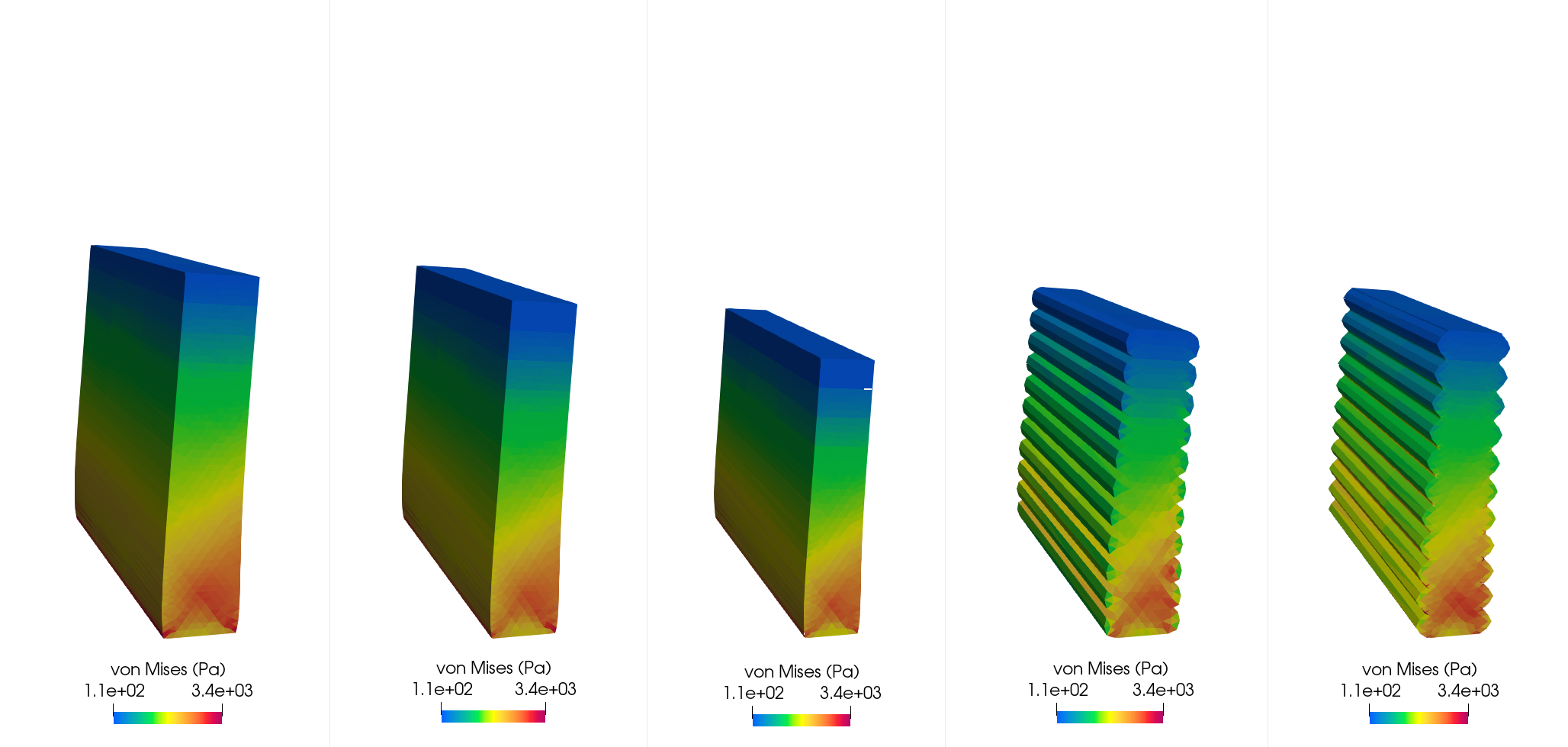}
    \caption{Material LM 30 (with density adjustment): Stress contour plot over deformed structure at point of failure using different filament representations (\textit{rect. width}, \textit{rect. adapted}, \textit{rect. contact}, \textit{ellipse}, \textit{points}). The deformation is scaled by a factor of 4.}
    \label{fig:validation_LM30}
\end{figure}

\begin{figure}[t]
    \centering
    \includegraphics[width=0.95\linewidth]{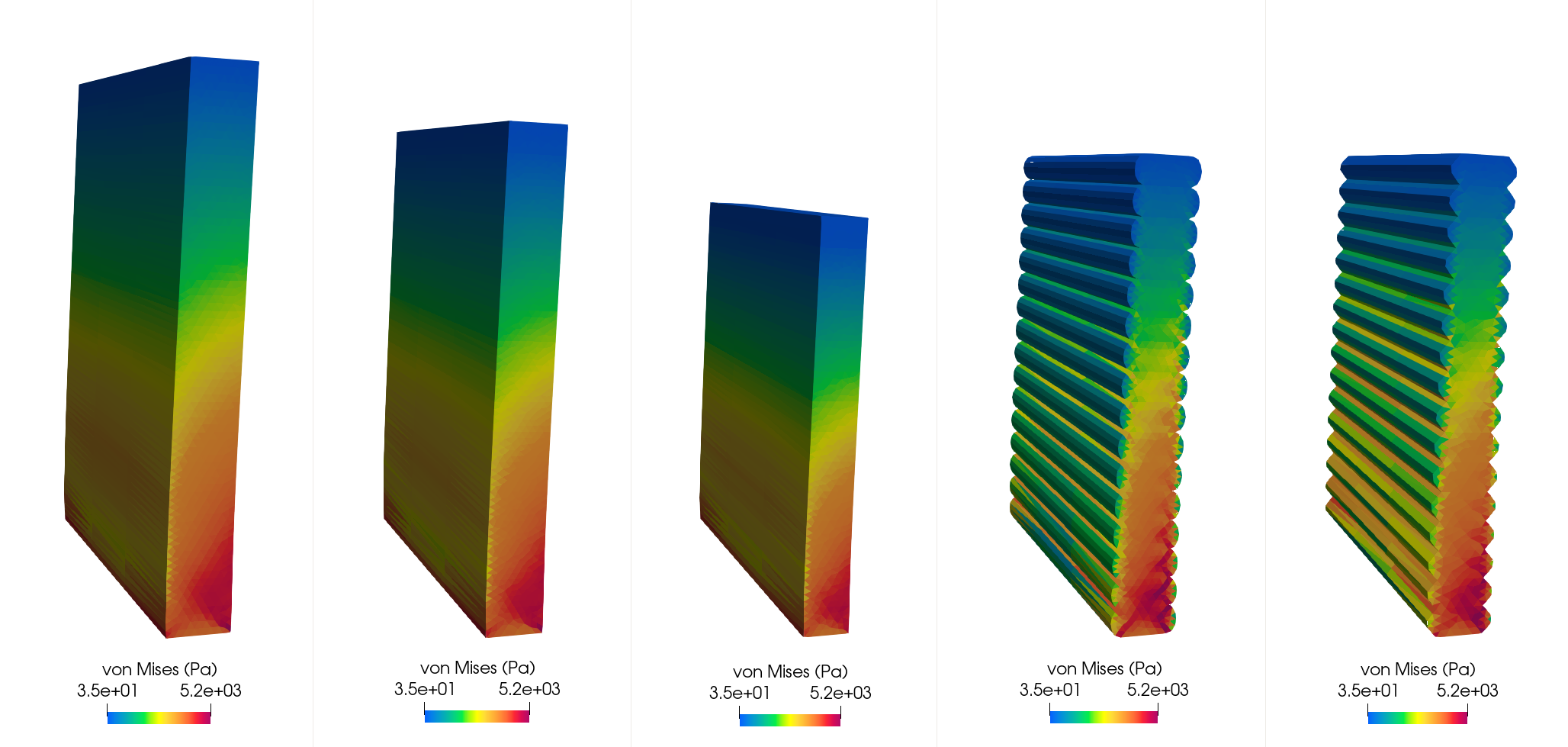}
    \caption{Material LSM 30 (with density adjustment): Stress contour plot over deformed structure at point of failure using different filament representations (\textit{rect. width}, \textit{rect. adapted}, \textit{rect. contact}, \textit{ellipse}, \textit{points}). The deformation is scaled by a factor of 4.}
    \label{fig:validation_LSM30}
\end{figure}

The influence of the filament representation is more pronounced for the stiffer material LSM~30, for which a larger number of layers can be printed before failure. Nevertheless, the \textit{ellipse} and \textit{points} representations show very good agreement with each other, whereas the \textit{rect. width} representation overestimates, and the \textit{rect. contact} representation underestimates the buildability. The \textit{rect. adapted} representation predicts failure at an intermediate level.  This observation is independent of the density adjustment. Figures~\ref{fig:validation_LM30} and~\ref{fig:validation_LSM30} illustrate the simulated failure states for the different filament representations, showing the observed trends of over- and underestimation. The contour plots show the von Mises stress distribution of the deformed structure at the point of failure computed with density adjustment.

\subsubsection{Influence of density adjustment}
As expected, the density adjustment does not affect the reference (\textit{points}) and the volume-preserving \textit{rect. adapted} representations. Furthermore, its influence is negligible for the softer material LM~30, which fails at a comparatively low number of layers. For the stiffer material LSM~30, differences of less than three layers are observed for the non-volume-preserving representations. Without density adjustment, the \textit{rect. width} representation predicts lower buildability, as the increased cross-sectional area results in a higher structural weight and, consequently, larger self-weight loading. Conversely, the \textit{rect. contact} representation predicts higher buildability because of its lower total mass. Since the volume of the \textit{ellipse} representation is very close to that of the \textit{points} representation, the influence of the density adjustment is negligible in this case. As a result, the overestimation associated with \textit{rect. width} and the underestimation associated with \textit{rect. contact} are partially compensated when no density adjustment is applied. To isolate the effect of the geometric representation, the density-adjusted set-up is adopted for all subsequent investigations.

Overall, good agreement is obtained between simulation and experiment. While this comparison does not constitute a complete validation of the framework, it demonstrates that the model reproduces the experimentally observed values and highlights the importance of the assumed filament cross-section for reliable prediction of buildability. This is sufficient for the subsequent investigation of the relative influence of different filament representations on buildability.

\subsection{Numerical investigation on the influence of filament shape on buildability: rectilinear walls}\label{sec:walls}
This section investigates the influence of process parameters and layer geometry on the buildability of rectilinear 3D-printed walls under different material behaviours. Specifically, we consider both changes in the physical filament shape resulting from different printing conditions (Sec. \ref{ss-3.2.1}) and those arising in the numerical model from different filament representations (Sec.~\ref{sec:filament}).

We first examine the influence of filament shape and its discretization on the elastic buckling of rectilinear walls, assuming a purely linear-elastic material model (Sec. \ref{ss-3.2.2}). This allows us to isolate the effects of process parameters and filament shape on buildability from those arising from material nonlinearities. Within the same linear-elastic framework, we then investigate the combined effects of filament-shape discretization and varying Young's modulus on buildability predictions (Sec. \ref{ss-3.2.3}) using a fixed printing setup. The numerical results are also compared with an analytical buckling approach, providing further validation of the numerical framework.

Finally, we investigate the buildability of rectilinear walls using a more realistic elasto-plastic material model (Sec. \ref{ss-3.2.4}). Specifically, two representative materials with distinct mechanical properties are considered: one characterised by a relatively low yield stress and pronounced plasticity, and the other by a higher yield stress and a more extended elastic regime. Together, these materials span the transition from buildability failure dominated by plastic yielding to failure governed primarily by elastic buckling.

\subsubsection{Process parameters and filament geometry prediction}\label{ss-3.2.1}
First, six representative printing conditions are defined to span a realistic range of operating regimes. The nozzle diameter is kept constant at $\phi_n = 20\,\mathrm{mm}$, while two nozzle heights are considered to capture distinct deposition behaviours. Namely, a smaller nozzle height $h_n = 10\,\mathrm{mm}$ is employed to achieve a more layer-pressing printing condition, while a larger one, $h_n = 20\,\mathrm{mm}$, is representative of a more free-flow deposition regime. In addition to the nozzle geometric parameters, the printing regime is also affected by the speed ratio $v^*=v_p/u_f$. For each nozzle height, three cases are generated by varying the print and inflow velocities to obtain $v^* = 0.5,\,1,\,2$. This results in six distinct combinations, identified as ID1--ID6, which systematically explore the coupled effects of nozzle geometry and kinematics; the corresponding parameters are summarized in Table~\ref{tab:3}.

Material properties in the fluid state are kept identical across all cases to isolate the effect of process parameters. Specifically, the density is fixed at $\rho = 2112.5\,\mathrm{kg/m^3}$, the plastic viscosity at $\mu = 15\,\mathrm{Pa \cdot s}$, and the yield stress is set to a representative value of $1300\,\mathrm{Pa}$, chosen as an intermediate value between the two considered in the validation (Section~\ref{sec:validation}).

\begin{table}[h]
\centering
\caption{Process parameters of the six different print scenarios ID1-ID6.}
\label{tab:3}
\begin{tabular}{@{} c ccccc @{}}
\toprule
& \textbf{ID} &
$\phi_n$ (mm) &
$h_n$ (mm) &
$v_p$ (mm/s) &
$u_f$ (mm/s) \\
\midrule
& ID1 & 20   & 10   & 20   & 40 \\
& ID2 & 20   & 10   & 20   & 20 \\
& ID3 & 20   & 10   & 40   & 20   \\
& ID4 & 20   & 20   & 20   & 40   \\
& ID5 & 20   & 20   & 20   & 20   \\
& ID6 & 20   & 20   & 40   & 20   \\
\bottomrule
\end{tabular}
\end{table}

\begin{figure}[h]
    \centering
    \includegraphics[width=0.9\linewidth]{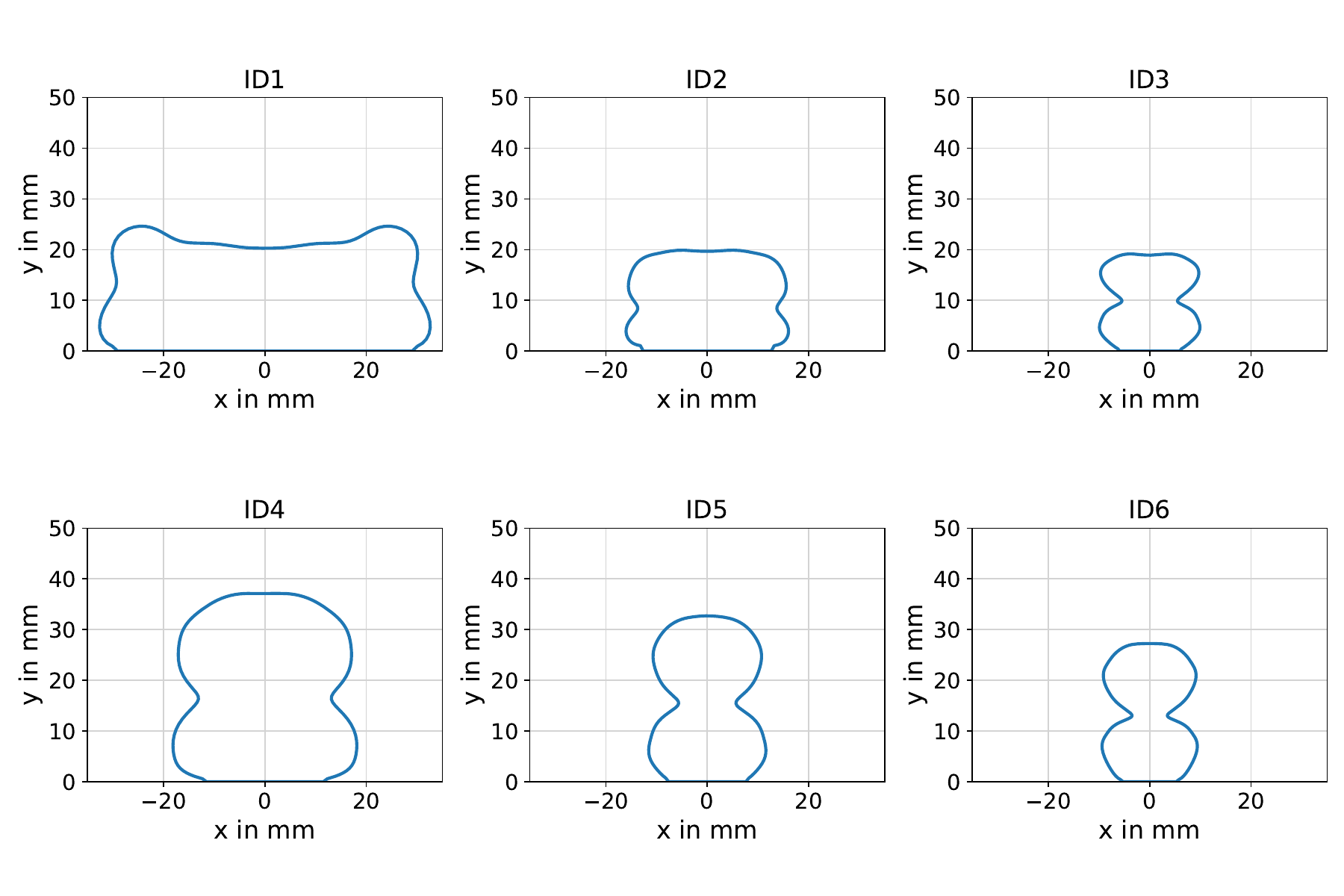}
    \caption{Two-layers cross-sectional geometries predicted for the six different print scenarios ID1-ID6.}
    \label{fig:3}
\end{figure}

Subsequently, filament geometries for all print scenarios are predicted using \textit{ShapeGen3DCP}. As illustrated in detail in \cite{rizzieri2026}, \textit{ShapeGen3DCP} provides fast and reliable predictions, with typical deviations of 1--10\% from experimental measurements. These deviations are comparable to the intrinsic uncertainties of the printing process and can therefore be considered acceptable for estimating the actual filament shape. For each condition, the cross-sectional profiles of two superimposed layers are computed and reported in Figure~\ref{fig:3}.

The generated cross-sections are also post-processed to extract the key geometric features reported in Table~\ref{tab:4}, namely the base layer width ($w_1$), the base layer height ($h_1$), and the interlayer contact length ($w_{c}$). From these quantities, the filament aspect ratio of the base layer ($w_1/h_1$) is computed, providing a direct measure of the layer spreading and the filament roundness ratio $w_1/w_{c}$ is defined to characterize the curvature of the sides of the layer's profile.

\begin{table}[h]
\centering
\caption{Post-processed geometrical features, roundness ratio and two-layers area.}
\label{tab:4}
\begin{tabular}{@{} c ccccccccc @{}}
\toprule
& \textbf{ID} &
$h_n$ ($mm$) &
$v^*$ ($-$) &
$w_1$ ($mm$) &
$w_{c}$ ($mm$) &
$h_1$ ($mm$) &
$w_1/h_1$ ($-$) &
$w_1/w_c$ ($-$) &
$A_{SG}$ ($mm^2$)\\
\midrule
& ID1 & 10 & 0.5 & 65.2    & 58.5  & 13.6 & 4.8 & 1.1  & 1337.2\\
& ID2 & 10 & 1 & 32.1    & 27.4  & 8.4 & 3.8 & 1.2  & 567.8 \\
& ID3 & 10 & 2 & 19.8    & 10.97 & 9.8 & 2.0 & 1.8  & 316.0 \\
& ID4 & 20 & 0.5 & 36.3    & 26.2 & 16.3 & 2.2  & 1.4  & 1145.8 \\
& ID5 & 20 & 1 & 23.1    & 11.3 & 15.4 & 1.5 & 2.0  & 605.3 \\
& ID6 & 20 & 2 & 18.8    & 7.0 & 13 & 1.44  &  2.7  & 407 \\
\bottomrule
\end{tabular}
\end{table}

From the filament cross-sectional results, different printing regimes can be identified. Cases ID1 and ID2 exhibit flatter and wider filaments, associated with high aspect ratios ($w_1/h_1 \gg 1$), which indicate pronounced lateral spreading and a roundness ratio close to unity ($w_1/w_c \simeq 1$), i.e., nearly rectangular profiles with limited curvature. These features reflect the outcome of a marked layer-pressing printing regime, associated with moderately high speed ratios and a small nozzle height.

In contrast, cases ID5 and ID6 produce rounder and taller filaments, with aspect ratios closer to unity ($w_1/h_1 \simeq 1$), reflecting reduced spreading and increased filament height. These cases also exhibit larger values of the roundness ratio ($w_1/w_c \gg 1$), which indicate more pronounced curvature. This behaviour is indicative of a free-flow deposition regime, in which the print speed is generally higher than the inflow speed, and the nozzle height is comparable to or greater than the nozzle diameter.

Cases ID3 and ID4 lie between the layer-pressing and free-flow deposition regimes, due to their intermediate values of aspect ratio and roundness ratio. In fact, for ID3, the effect of high print speed is balanced by the reduced nozzle height, whereas for ID4, the opposite holds.

Table~\ref{tab:4} also reports the predicted cross-sectional areas. The influence of the speed ratio on filament geometry is clearly observed: within each group of three cases (i.e., at fixed nozzle height), the cross-sectional area decreases as the speed ratio increases. This trend reflects the inverse relationship between the deposited material volume per unit length and the speed ratio, in agreement with mass conservation.

\subsubsection{Influence of filament shape on elastic buckling}\label{ss-3.2.2}

Buckling simulations are performed for the six printing scenarios defined in Table~\ref{tab:3} and characterized in Table~\ref{tab:4}, considering $200$~mm long wall structures with a fixed base (see Section~\ref{sec:simulation}) and varying filament representations (\textit{rect. width, rect. contact, rect. adapted, ellipse, point} as defined in Section~\ref{sec:filament}).

To isolate pure elastic buckling effects, an elastic material model is considered, using for simplicity the elastic parameters obtained from the calibration of material LM~30 (see Table~\ref{tab:material_parameters}). The Young's modulus is rounded to $990000$~Pa, with a rate of increase over time of $22.11$~Pa/s, and a Poisson's ratio of $0.3$.

As in the validation study, an imperfection magnitude of $\alpha = 0.00025$ and a load step of $0.25$~s are applied. To ensure a fair comparison between the different geometrical representations, the density adjustment setup studied in Sec.~\ref{sec:validation} is used throughout the following analyses to preserve the total mass and, consequently, the dead load.

\begin{figure}[t]
    \centering
    \subfloat[Nozzle height $h_n=10$~mm]{\includegraphics[width=0.43\linewidth]{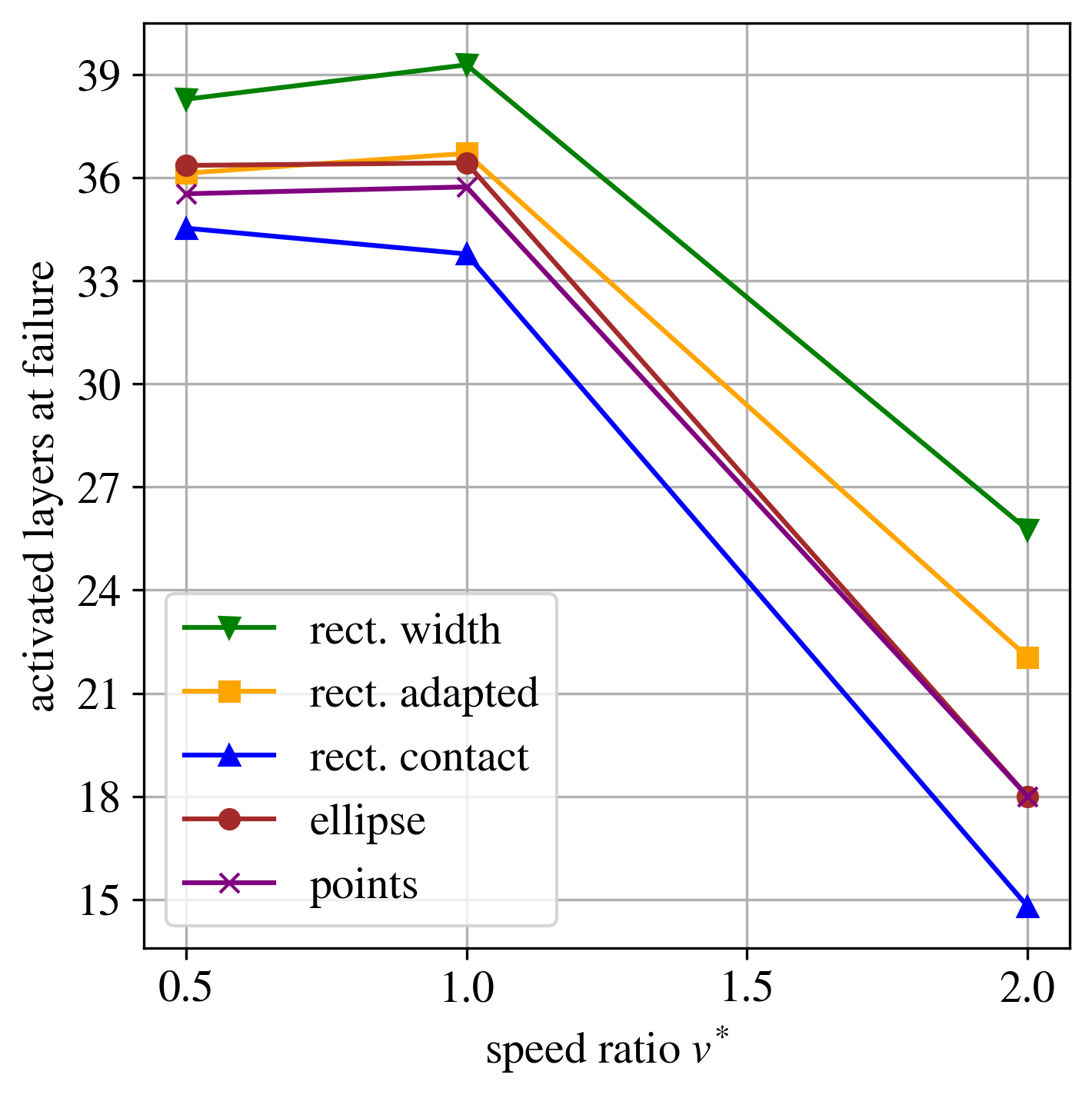}} %
    \subfloat[Nozzle height $h_n=20$~mm]{\includegraphics[width=0.43\linewidth]{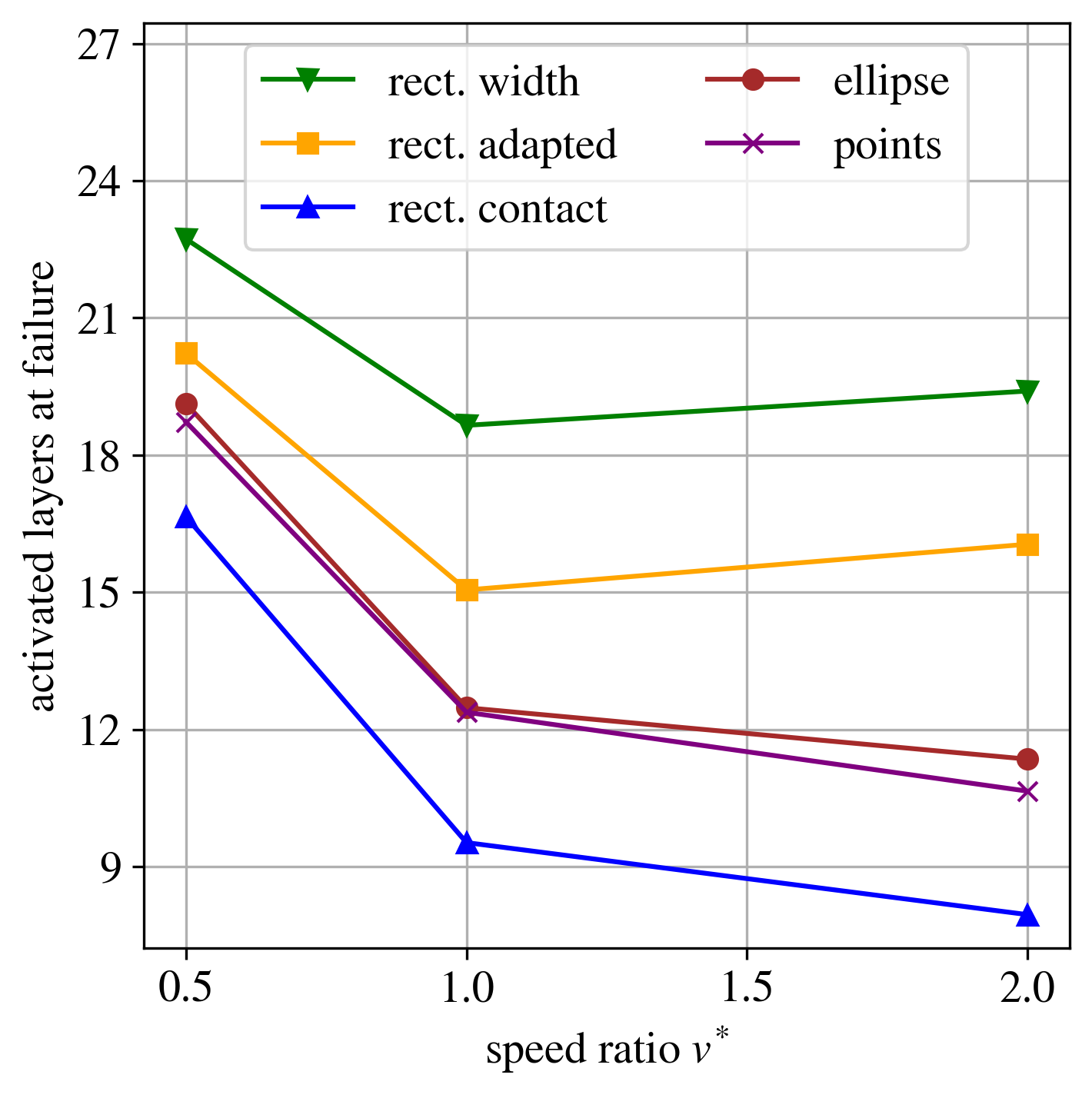}}
    \caption{Elastic study: Number of activated layers at failure predicted for the different filament representations and printing scenarios listed in Table~\ref{tab:3}.}
    \label{fig:elastic_study}
\end{figure}

Figure~\ref{fig:elastic_study} reports the number of activated layers at failure, as defined in eq.~\eqref{eq:num_activated_layers}, for all printing scenarios. The results are shown as a function of the speed ratio, separately for nozzle heights of $10$~mm (ID1--ID3, left) and $20$~mm (ID4--ID6, right). 
Consistent with the validation study, the point-based and elliptical filament representations lead to nearly identical buildability predictions. In contrast, \textit{rect. contact} systematically underestimates the number of buildable layers, whereas \textit{rect. width} systematically overestimates it. The \textit{rect. adapted} representation also overestimates the buildability, but to a lesser extent than \textit{rect. width}. Note, a similar number of activated layers for different printing scenarios (e.g., for ID1 and ID2) leads to different absolute heights at failure since the layer heights differ. 

\begin{figure}[t]
    \centering
    \includegraphics[width=0.425\linewidth]{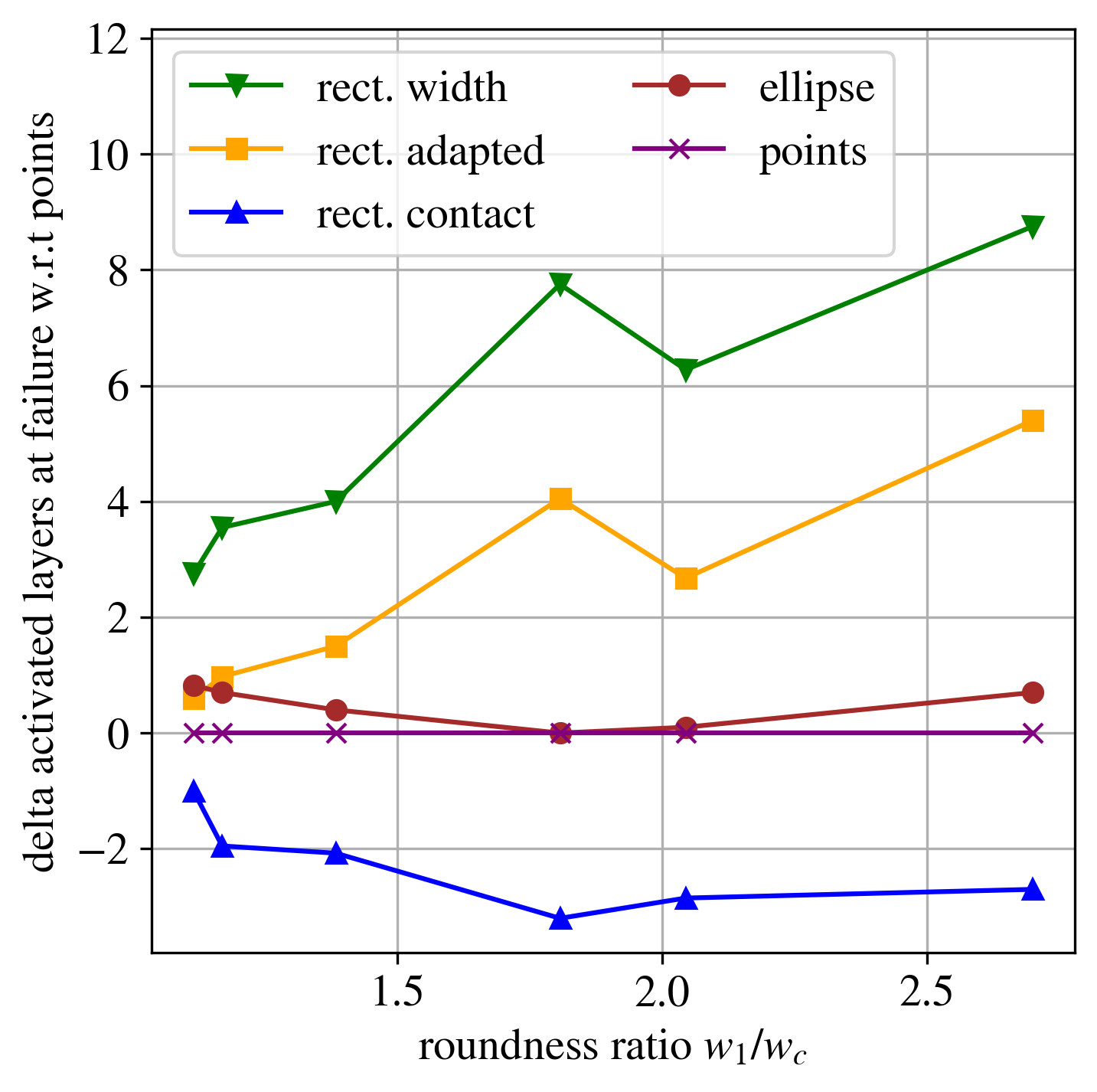}
    \caption{Elastic study: Absolute difference in the number of activated layers at failure with respect to the \textit{points} representation for the six printing scenarios listed in Table~\ref{tab:3} shown as a function of the roundness ratio.}
    \label{fig:elastic_study_roundness}
    
\end{figure}

A clear influence of the process parameters can also be observed. For cases with pronounced layer pressing (Figure~\ref{fig:elastic_study}a, speed ratios $0.5$ and $1.0$, corresponding to ID1 and ID2), the effect of the filament representation is less pronounced than for the free-flow cases (Figure~\ref{fig:elastic_study}b, speed ratios $1.0$ and $2.0$, corresponding to ID5 and ID6), which exhibit more rounded filament geometries. To quantify this trend, Figure~\ref{fig:elastic_study_roundness} shows the absolute difference in the number of activated layers at failure with respect to the \textit{points} representation for all six printing scenarios listed in Table~\ref{tab:4}. For each representation, this difference is computed as the number of activated layers at failure of the respective representation minus that obtained with \textit{points}. With increasing roundness ratio, the deviation from \textit{points} increases. For \textit{rect. contact}, the difference decreases from approximately $-1$ to about $-3$ layers, indicating an increasing underestimation of buildability. For \textit{rect. adapted}, the difference increases from about $1$ to $5$ layers, while for \textit{rect. width} it increases from about $3$ to approximately $9$ layers, indicating an increasing overestimation. In contrast, the difference between \textit{ellipse} and \textit{points} remains below one layer for all cases considered, showing that \textit{ellipse} captures the influence of the filament shape very accurately, independently of the roundness ratio.

\subsubsection{Combined influence of filament shape discretization and Young's modulus on elastic buckling}\label{ss-3.2.3}
A second study is conducted to specifically investigate the combined influence of Young’s modulus and filament-shape representation on the buildability of rectilinear walls. For this purpose, printing scenario ID3 ($\Phi = 2$ and $h_n = 10$~mm) is selected, as it produces a pronounced, rounded filament geometry and, consequently, exhibits a marked sensitivity to the adopted filament representation, as observed in the previous comparison. The material is still assumed to behave linearly elastically, while the Young’s modulus is varied from $61.875$~kPa to $1980$~kPa, thereby covering the broad range of values reported in the literature. The rate of increase of Young’s modulus over time is kept unchanged, as are all other simulation settings, consistently with the elastic simulations described in Section \ref{ss-3.2.2}.

Following the bifurcation analysis from Suiker~\cite{suiker2018}, the analytical expression for the critical wall height is used to compute a reference value for comparison:
\begin{equation}\label{eq:critical_height}
    l_{crit} = 1.98635 \left(\frac{D}{\rho g w_i}\right)^{1/3},
\end{equation}
with $D=\frac{E w_i^3}{12 (1-\nu)^2}$. $w_i$ denotes the wall width associated with the selected rectangle-based filament representation ($w_1$, $w_c$, or $w_{ad}$). For the analytical solution, a constant Young's modulus is assumed, i.e., no evolution of the stiffness is considered. The gravitational acceleration is set to $g=9.81$~m/s$^2$, and the density $\rho$ used in the corresponding structural simulation is adopted for the calculation of the analytical critical height. The corresponding number of layers is computed by dividing $l_{crit}$ by the layer height $h_1$.

\begin{figure}[h]
    \centering
    \includegraphics[width=0.9\linewidth]{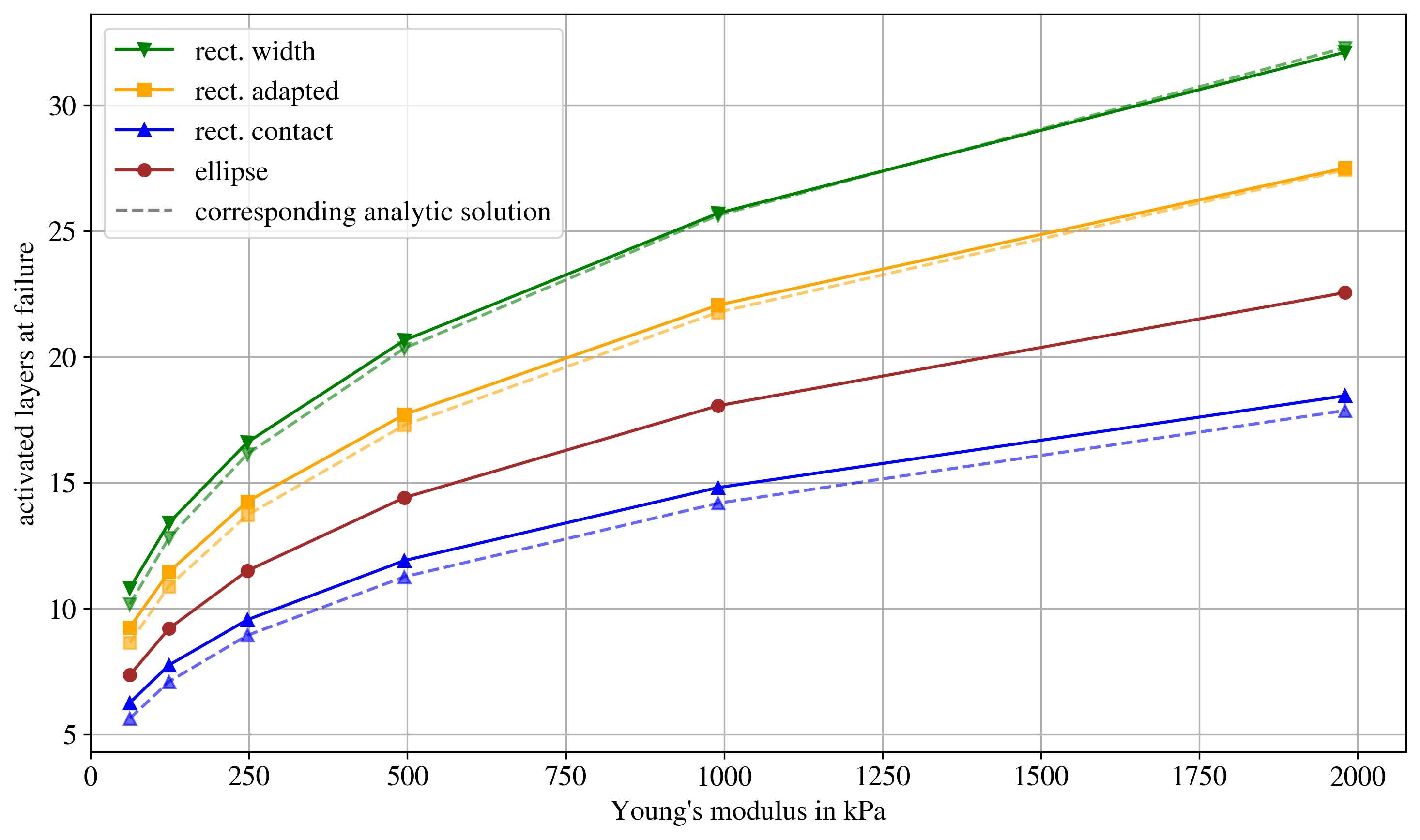}
    \caption{Elastic ID3 case: Number of activated layers at failure for varying Young's moduli. The dashed lines indicate the analytical prediction obtained from eq.~\eqref{eq:critical_height}, expressed as $l_{crit}/h_1$, for the rectangle-based filament representations. For the \textit{ellipse} case, only the simulation result is shown.}
    \label{fig:youngs_modulus}
\end{figure}

Figure~\ref{fig:youngs_modulus} showcases the simulated number of activated layers at failure for the different Young’s moduli considered, together with the predictions obtained from the analytical formula. First, it can be observed that, for all rectangle-based filament representations, the simulation results are in good agreement with the analytical solution over the entire range of Young’s moduli investigated. The small deviations between the numerical and analytical results can be attributed to differences in the underlying assumptions, including the stopping criterion and the reference configuration adopted. Overall, these results further support the validity of the numerical framework in reliably capturing the governing elastic buckling behaviour.
 

As expected, the maximum number of buildable layers decreases with decreasing Young's modulus, as a reduction in material stiffness lowers the structural resistance to elastic instability and buckling under self-weight. Consistent with the previous study, \textit{rect. contact} leads to lower buildability predictions, whereas \textit{rect. width} and \textit{rect. adapted} yield higher values. The \textit{ellipse} representation, which is only evaluated numerically since the analytical expression in Eq.~\eqref{eq:critical_height} is derived exclusively for rectangular cross-sections, again lies between the rectangle-based approaches.

\begin{table}[h]
\centering
\caption{Elastic ID3 case: Percentage changes in the number of activated layers for varying Young's moduli [kPa] and filament representations with respect to \textit{ellipse}.}
\label{tab:actlayers_youngs_modulus_relative}
\begin{tabular}{lrrrrrr}
\toprule
E in kPa & 61.875 & 123.75 & 247.5 & 495 & 990 & 1980 \\
\midrule
\textit{rect. width} & 46.94 & 45.65 & 44.35 & 43.40 & 42.38 & 42.35 \\
\textit{rect. adapted} & 25.85 & 24.46 & 23.91 & 22.92 & 22.16 & 21.95 \\
\textit{rect. contact} & -14.97 & -15.76 & -16.96 & -17.36 & -18.01 & -18.18 \\
\bottomrule
\end{tabular}
\end{table}

To quantify the influence of the filament representation, Table~\ref{tab:actlayers_youngs_modulus_relative} reports the percentage change in the number of activated layers at failure with respect to the corresponding value obtained using the \textit{ellipse} representation. The relative deviations are nearly independent of Young's modulus. For smaller Young's moduli, the lower number of buildable layers results in a higher sensitivity of the percentage values. As the Young's modulus increases, the relative deviations converge and approach nearly constant values. This observation suggests that the influence of the filament representation can be generalized across elastic materials. 
For the wall simulations considered here, the \textit{rect. contact} representation underestimates the buildability by up to $18\,\%$, whereas \textit{rect. adapted} and \textit{rect. width} overestimate it by up to $26\,\%$ and $47\,\%$, respectively, relative to the \textit{ellipse} representation.


\subsubsection{Influence of filament shape on elasto-plastic buckling}\label{ss-3.2.4}

The influence of filament shape and its representation on buildability is further investigated in the case of elasto-plastic material behaviour, specifically using the two material models LM~30 and LSM~30 introduced in Section~\ref{sec:validation}. The six printing scenarios defined in Table~\ref{tab:3} are thus again reproduced, with all five filament representations considered for each scenario. The simulations are performed on $200$~mm-long wall structures, using the same boundary conditions, numerical settings, and assumptions adopted in the elastic studies. These include fixed displacements at the base, the same imperfection magnitude, identical temporal and spatial discretizations, and density adjustment to preserve the total mass, and hence the dead load, irrespective of the adopted layer-shape representation.

\begin{figure}[h]
    \centering
    \subfloat[Nozzle height $h_n=10$~mm]{\includegraphics[width=0.45\linewidth]{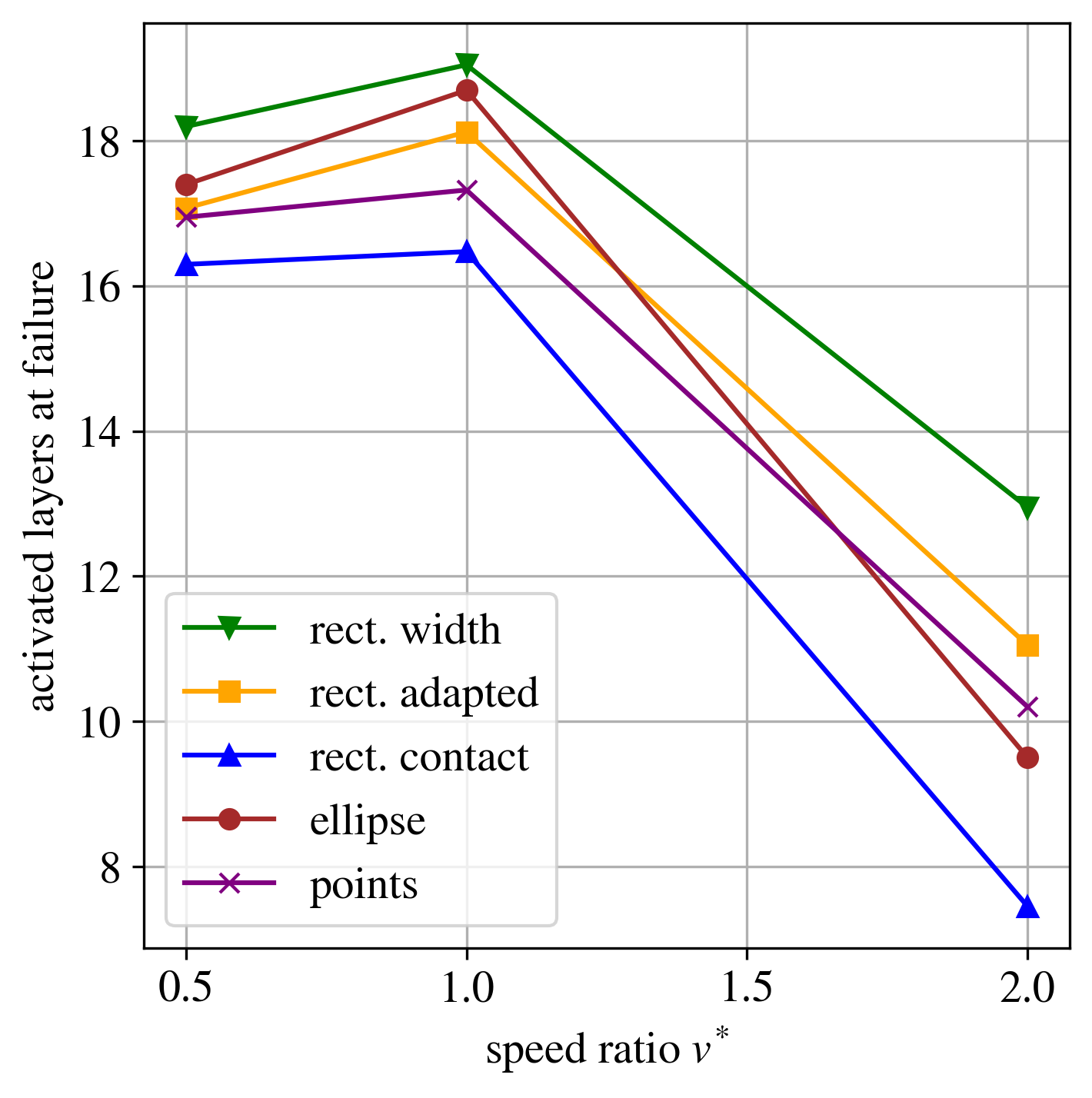}} 
    \subfloat[Nozzle height $h_n=20$~mm]{\includegraphics[width=0.45\linewidth]{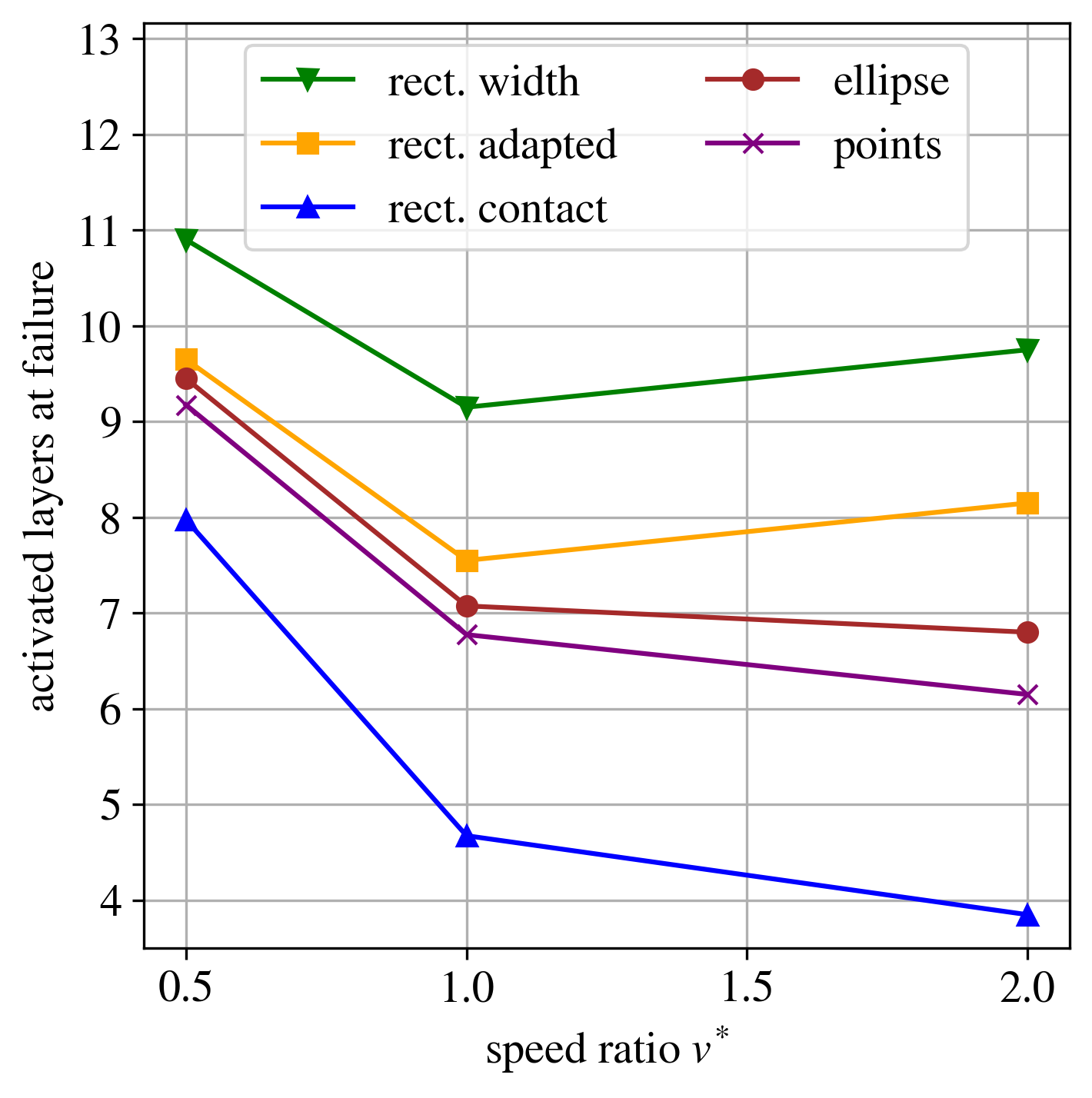}}
    \caption{Plastic study LM 30: Comparison of the number of layers at failure using different filament representations for the different printing cases.}
    \label{fig:plastic_study_LM30}
\end{figure}

\begin{figure}[h]
    \centering
   \subfloat[Nozzle height $h_n=10$~mm]{\includegraphics[width=0.45\linewidth]{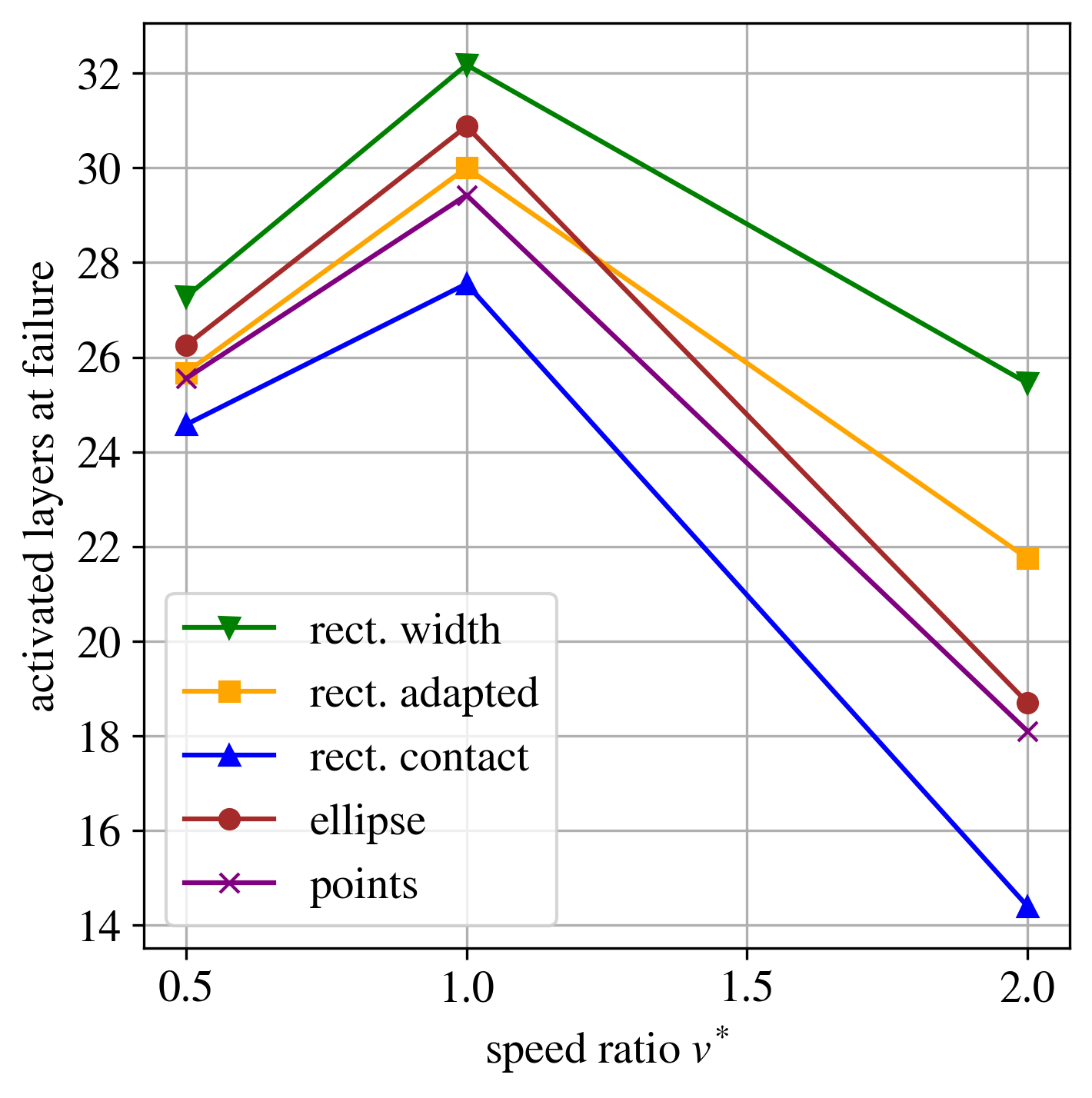}} %
    \subfloat[Nozzle height $h_n=20$~mm]{\includegraphics[width=0.45\linewidth]{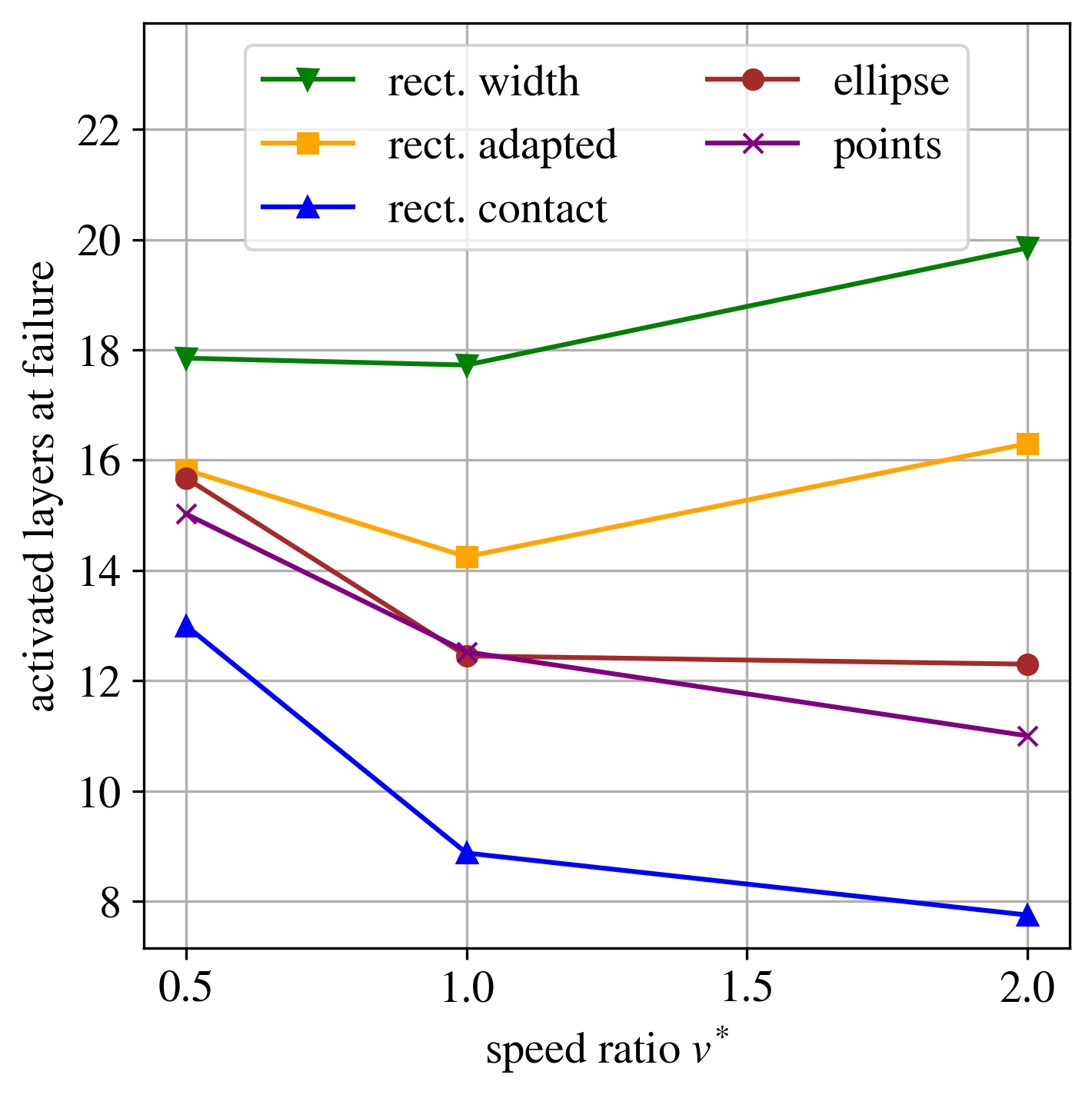}}
    \caption{Plastic study LSM 30: Comparison of the number of layers at failure using different filament representations for the different printing cases.}
    \label{fig:plastic_study_LSM30}
\end{figure}

\begin{figure}[t]
    \centering
   \subfloat[LM~30]{\includegraphics[width=0.45\linewidth]{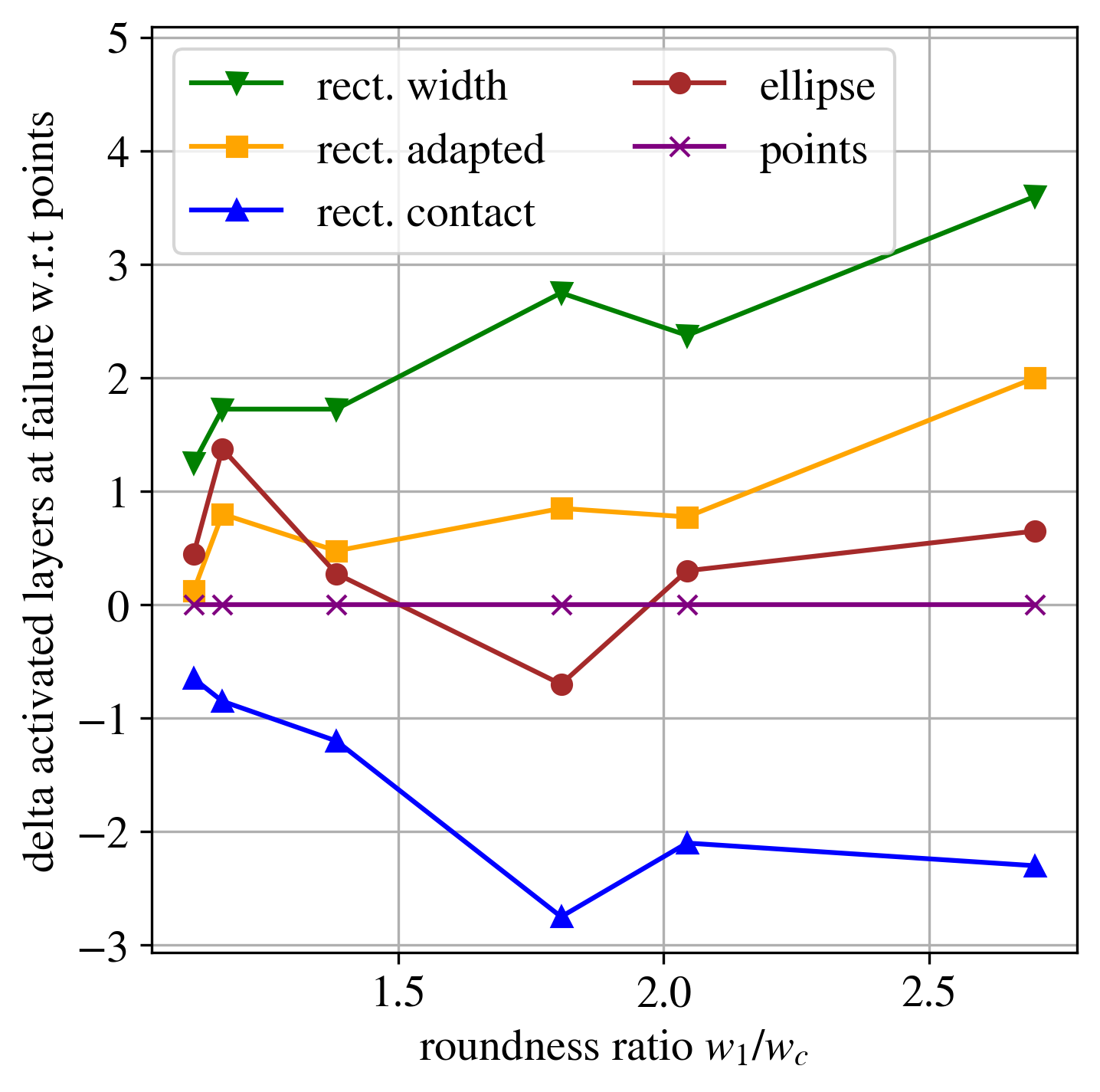}} %
    \subfloat[LSM~30]{\includegraphics[width=0.45\linewidth]{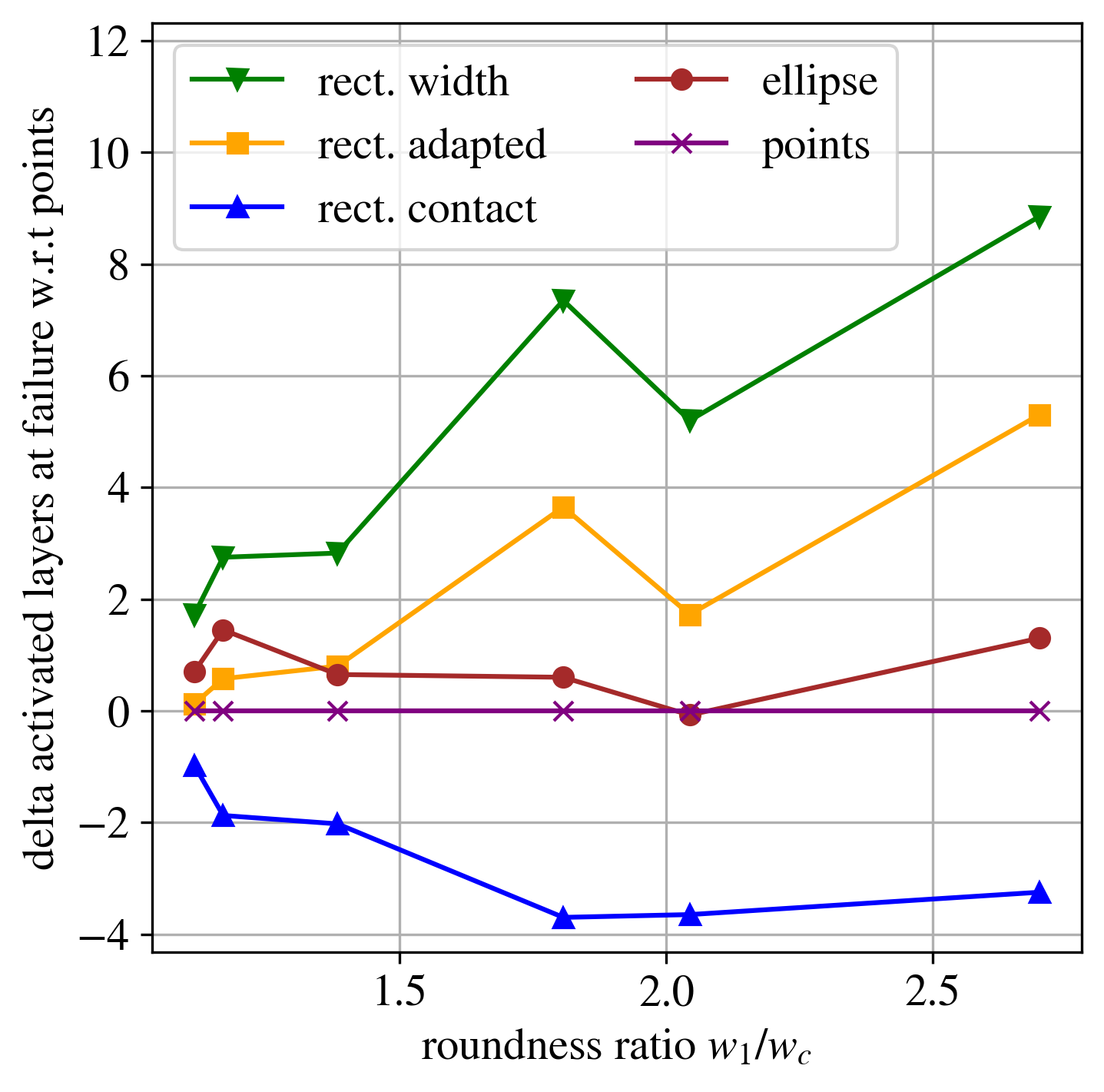}}
    \caption{Plastic study: Absolute difference in the number of activated layers at failure with respect to the \textit{points} representation for the six printing scenarios listed in Table~\ref{tab:3}, shown as a function of the roundness ratio for both materials.}
    \label{fig:plastic_study_roundness}
\end{figure}

Figures~\ref{fig:plastic_study_LM30} and~\ref{fig:plastic_study_LSM30} show the number of activated layers at failure for LM~30 and LSM~30 for each layer shape, respectively. Again, the results are shown as a function of the speed ratio for all printing scenarios, with a nozzle height of $10$~mm on the left and $20$~mm on the right. Compared to the purely elastic case, the absolute number of activated layers at failure is reduced for both elasto-plastic materials, reflecting the earlier onset of collapse due to plasticity. Due to its higher yield stress and stiffer elastic response, LSM~30 consistently allows more layers to be built before elasto-plastic buckling than LM~30. For example, in print scenario ID2, the wall printed with LSM~30 fails within the 30rd layer, whereas using LM~30 it already fails within the 18th layer. In the pure elastic case, 36 layers were reached with similar elastic properties to LM~30.

The sensitivity to the filament representation follows the same trend as in the elastic case, being more pronounced for free-flow scenarios. The absolute values in the number of layers depend on the material properties and differ from the elastic study. For that, Figure~\ref{fig:plastic_study_roundness} reports the absolute difference in the number of activated layers at failure with respect to the \textit{points} representation as a function of the roundness ratio for both materials. As in the elastic case, the influence of the filament representation increases for more rounded filaments, corresponding to reduced layer pressing. For LM~30, the absolute deviations remain relatively small, with \textit{rect. width} overestimates the buildability by less than four layers. For LSM~30, the absolute deviations are larger and reach magnitudes comparable to the elastic study, with overestimations of up to approximately nine layers.

Overall, these results demonstrate that, when plastic yielding occurs, plasticity reduces the absolute buildability due to elasto-plastic buckling mechanisms but does not alter the fundamental influence of the filament representation. In particular, the \textit{ellipse} approximation fits the most realistic \textit{points} representation very well, whereas rectangle-based representations can lead to significant over- or underestimation of buildability, especially for more rounded filament geometries.


\newpage
\subsubsection{Comparative influence of filament shape and discretization on buildability
}
\begin{figure}[h]
    \centering
    \includegraphics[width=0.9\linewidth]{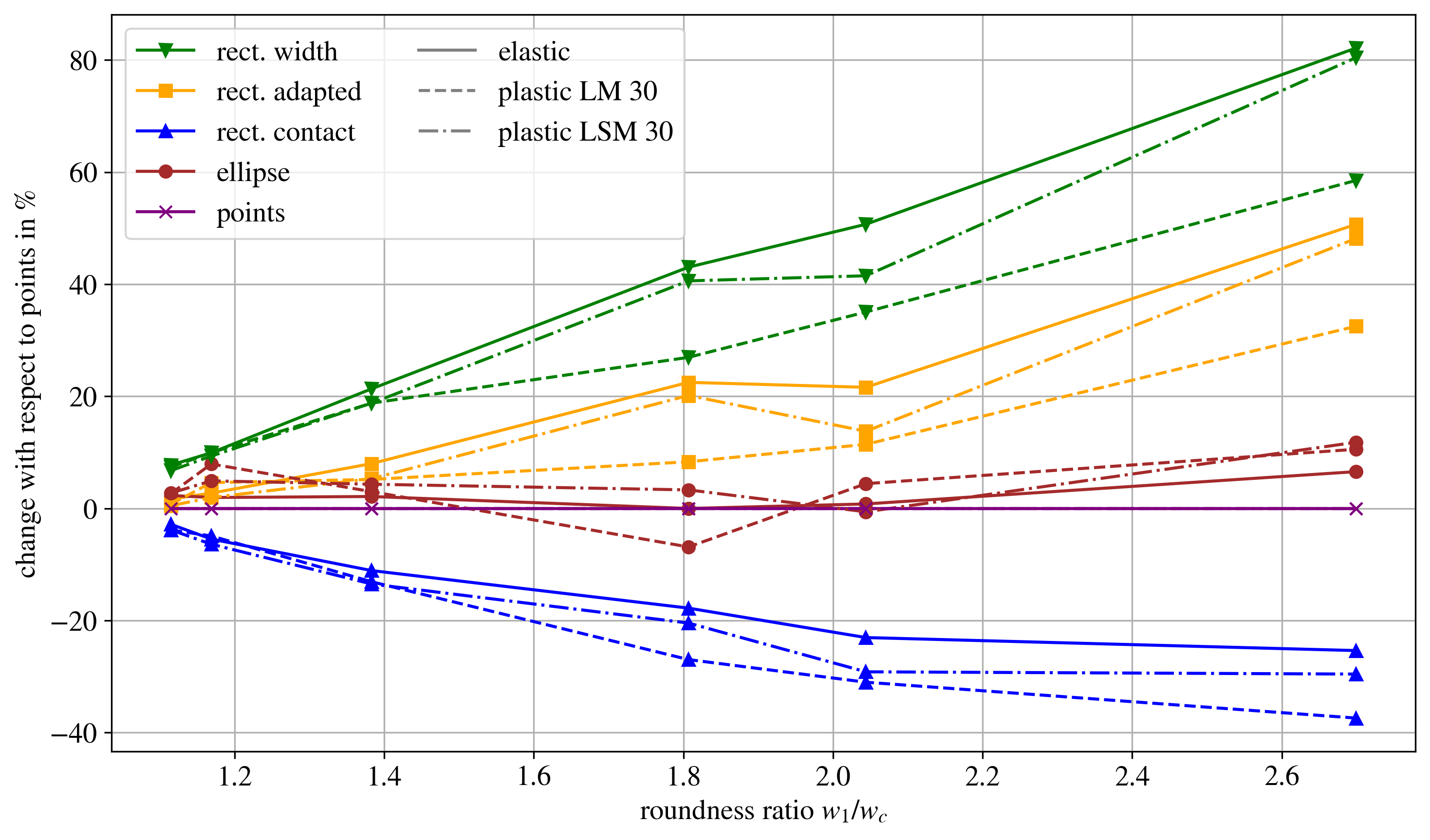}
    \caption{Change of number of activated layers with respect to the \textit{points} shape in percentage for the different cases: elastic (solid), plastic LM 30 (dashed), and plastic LSM (dash-dotted) as a function of roundness ratio.}
    \label{fig:relative_changes1}
\end{figure}

To enable a direct comparison across all investigated material models (purely elastic as well as elasto-plastic LM~30 and LSM~30), Figure~\ref{fig:relative_changes1} presents the percentage change in the number of activated layers at failure with respect to the corresponding value obtained with the \textit{points} representation, which is taken as the most realistic reference, over the roundness ratio. Although the absolute number of activated layers differs substantially between the elastic and elasto-plastic cases, as discussed in the preceding subsections, the relative changes with respect to the \textit{points} representation are remarkably similar for all materials.

For all material models, consistent trends are observed with respect to the filament representation and the roundness ratio. The influence of the filament shape increases for more rounded filament geometries, corresponding to reduced layer pressing and increased free-flow. \textit{Rect. width} and \textit{rect. adapted} approximations of the layer shape tend to overestimate the buildability, with relative increases of up to approximately $80\%$. In contrast, the \textit{rect. contact} approach underestimates the number of activated layers by up to about $30\%$. These relative deviations are largely independent of whether elastic or elasto-plastic material behavior is considered.
Furthermore, the elastic study with varying Young's modulus for case ID3 (Table~\ref{tab:actlayers_youngs_modulus_relative}) shows that the relative over- and underestimation introduced by the different filament representations remains largely unchanged over a wide range of Young's moduli. This observation complements the results obtained for the different material models and suggests that the influence of the filament representation is primarily governed by geometric rather than material effects.
Note that these results are based on the density-adjusted set-up. Without density adjustment, the observed over- and underestimation of the non-volume-preserving representations, especially \textit{rect. width} and \textit{rect. contact}, can be reduced (see Sect.~\ref{sec:validation}).

\begin{figure}[h]
    \centering
    \includegraphics[width=0.45\linewidth]{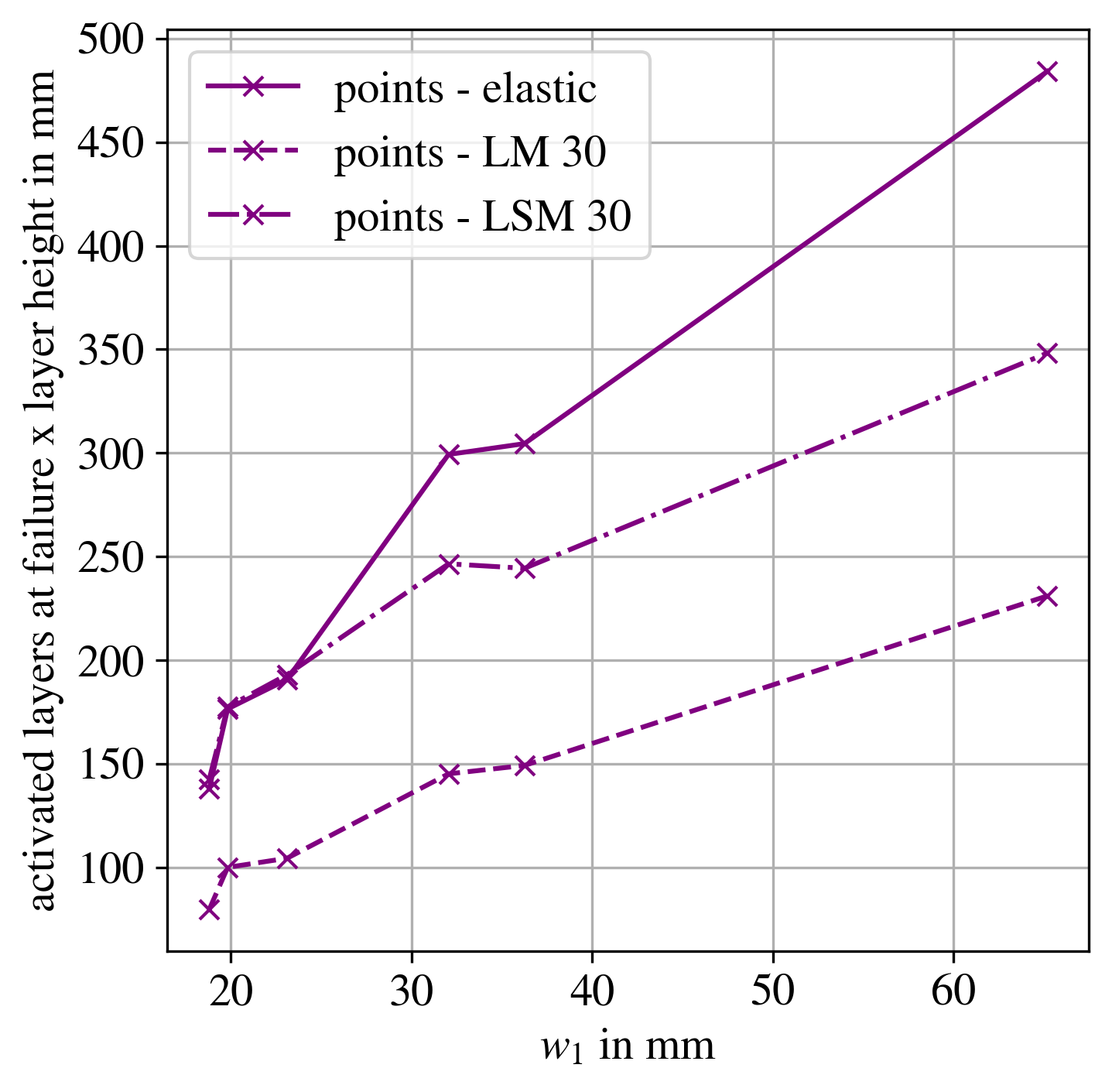}
    \caption{Wall height at failure (number of activated layers $\times$ layer height) for the \textit{points} representation and different materials: elastic (solid), plastic LM 30 (dashed), and plastic LSM (dash-dotted) as a function of the layer width $w_1$.}
    \label{fig:relative_changes2}
\end{figure}

In addition to filament shape effects, Figure~\ref{fig:relative_changes2} highlights the influence of geometric slenderness described by the layer width $w_1$ (the wall length is the same in all cases) on the absolute buildability. The wall height at failure is computed as the number of activated layers at failure multiplied by the layer height to account for the varying heights. The possible wall height increases with increasing layer width, i.e., for broader filament profiles. This trend is consistent across all material models and filament representations. For clarity, only the \textit{points} representation is shown in this comparison, as it provides the most realistic description of the filament geometry. The results confirm that less slender profiles allow higher total heights before elastic and elasto-plastic buckling. In general, plastic yielding reduces buildability compared to purely elastic material behavior. 


\clearpage
\subsection{Numerical investigation on the influence of filament shape on buildability: cylinders}\label{sec:cylinder}
Extending the investigation from rectilinear wall prints, the proposed framework is applied to study the influence of filament representations on the buildability of cylindrical structures with radii of $50$ and $150$~mm. 

The objective is to assess whether the trends identified for rectilinear walls in Sect.~\ref{sec:walls} also apply to cylindrical geometries. The analysis is intended as a qualitative assessment without validation with experimental data.
For that purpose, the printing scenario ID3 is selected, resulting in rounded filament profiles. The buildability simulations are performed using the softer elasto-plastic material LM~30. A larger imperfection amplitude of $\alpha = 0.005$ is prescribed to promote buckling in the $x$-direction. Under these conditions, a symmetric failure mode with respect to the x-axis develops, allowing the simulations to be carried out on one half of the cylinder using symmetric boundary conditions. The load step is set to $\Delta t = 2$~s leading to feasible computational costs, and a mesh size comparable to that used in the wall simulations is selected. As in the previous investigations, the density-adjusted set-up is adopted.

Table~\ref{tab:numlayers_cylinder} reports the resulting number of activated layers at the time of failure, defined as the point at which the out-of-plane deformation reaches one layer width (see Sect.~\ref{sec:simulation}). Since a close agreement between the \textit{points} and \textit{ellipse} representations was observed for rectilinear walls in Section~\ref{sec:walls}, the \textit{ellipse} representation is adopted here for simplicity as the most realistic reference. The same trends identified in the previous investigations are also observed for cylindrical geometries. The \textit{rect. width} shape leads to a pronounced overestimation of the buildability, which is reduced when using the \textit{rect. adapted} approach. In contrast, \textit{rect. contact} results in an underestimation with respect to the \textit{ellipse} case.  

\begin{table}[h]
\centering
\caption{Comparison of activated layers at failure for cylinders with two different radii and varying filament representations.}
\label{tab:numlayers_cylinder}
\begin{tabular}{lrrrr}
\toprule
 & \textit{rect. width} & \textit{rect. adapted} & \textit{rect. contact} & \textit{ellipse} \\
radius [mm] &  &  &  &  \\
\midrule
50 & 29.25 & 27.00 & 21.75 & 26.50 \\
150 & 41.58  &  33.17 & 19.17 & 30.83 \\
\bottomrule
\end{tabular}
\end{table}

\begin{figure}[b]
    \centering
    \includegraphics[width=0.83\linewidth]{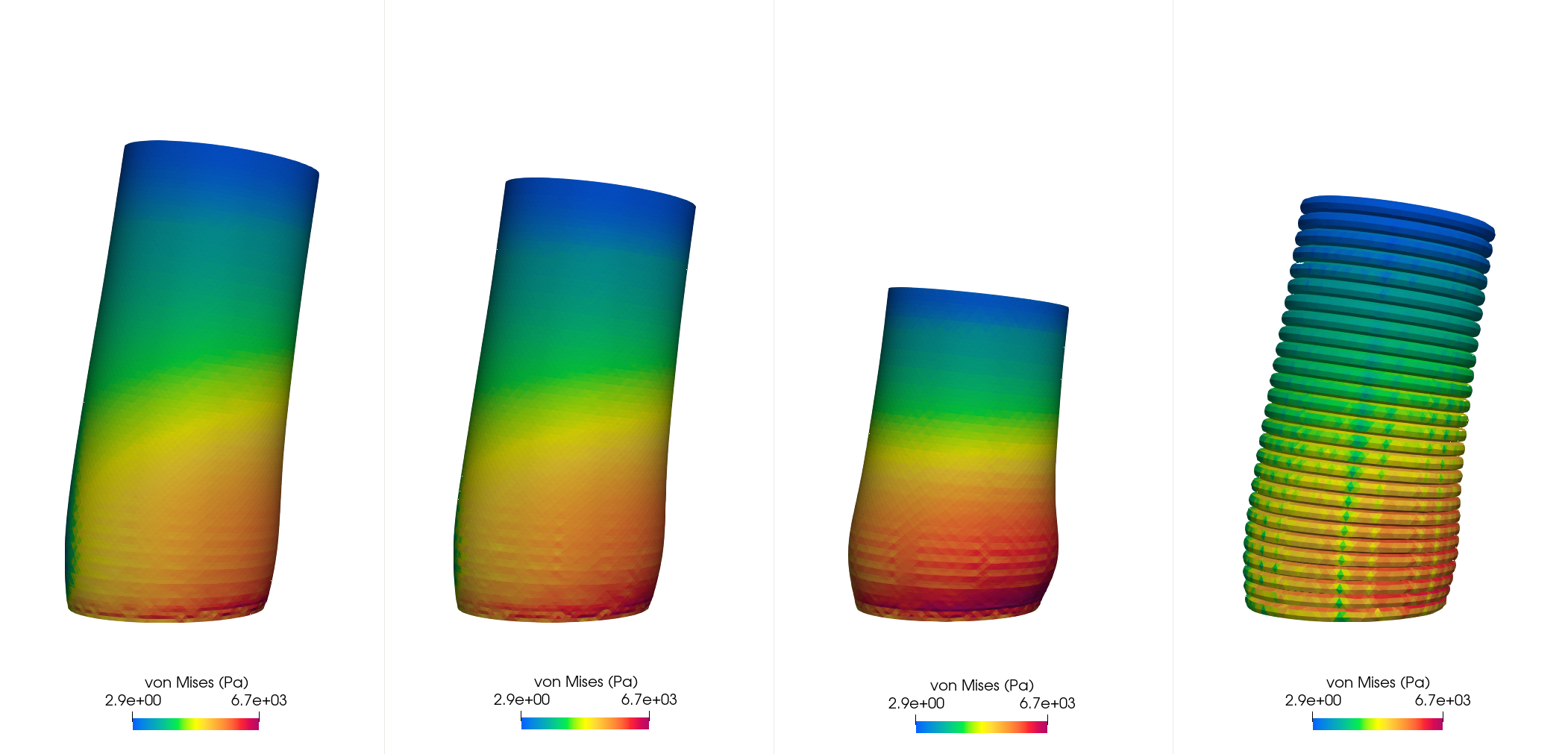}
    \caption{Cylinder radius $50$~mm: Stress contour plots over deformed structure at point of failure using different filament representations (\textit{rect. width}, \textit{rect. adapted}, \textit{rect. contact}, \textit{ellipse}). The deformation is scaled by a factor of 2.}
    \label{fig:cylinder_r50}
\end{figure}

\begin{figure}[h]
    \centering
    \includegraphics[width=0.83\linewidth]{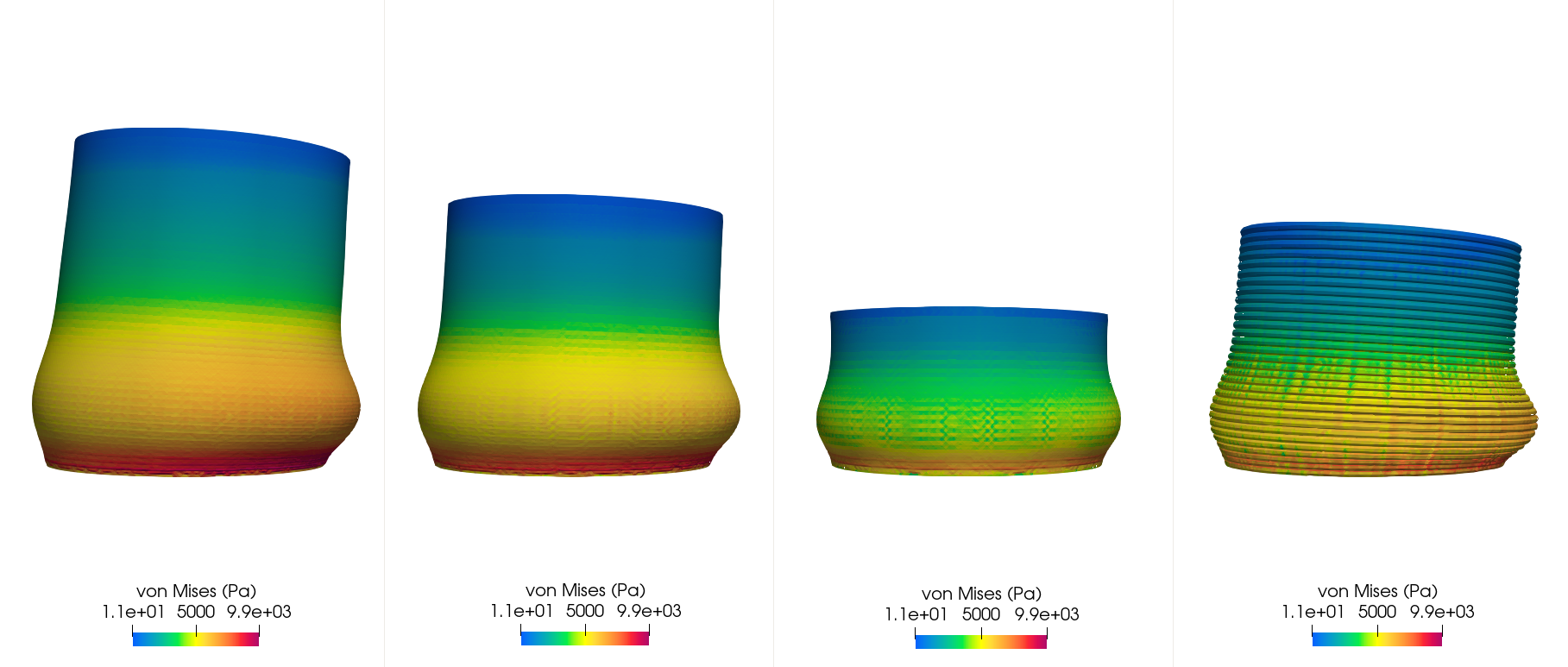}
    \caption{Cylinder radius $150$~mm: Stress contour plots over deformed structure at point of failure using different filament representations (\textit{rect. width}, \textit{rect. adapted}, \textit{rect. contact}, \textit{ellipse}). The deformation is scaled by a factor of 2.}
    \label{fig:cylinder_r150}
\end{figure}

Figure~\ref{fig:cylinder_r50} and Figure~\ref{fig:cylinder_r150} show the deformed configurations of the cylinders at the point of failure. Due to the different radii, two distinct failure mechanisms are observed, which influence the relative over- and underestimation rates with respect to the \textit{ellipse} representation. The more slender cylinder with $r = 50$~mm fails by global buckling, similar to the rectilinear wall case. In contrast, for the larger cylinder with $r = 150$~mm, failure is characterized by pronounced bulging in the upper region, which exceeds the prescribed out-of-plane deformation limit. In both cases, the failure mode itself is not altered by the choice of filament representation, confirming that this is governed by the global geometry. Nevertheless, for the larger cylinder radius, the magnitude of over- and underestimation relative to the \textit{ellipse} representation increases significantly. While for $r = 50$~mm the deviation ranges between approximately $-18\%$ and $+10\%$, it increases to about $-38\%$ and $+35\%$ for $r = 150$~mm.

\section{Discussion: results summary and simulation guidelines}\label{sec:4}
This section discusses the main findings of the study and derives general indications and guidelines that can support both designers and numerical modellers. In particular, the results provide insight into the influence of filament cross-sectional geometry on the buildability of thin-walled 3D-printed structures and on the sensitivity of buildability predictions to different geometric discretization choices adopted in numerical simulations.
\begin{itemize}
    \item The results demonstrate that the filament cross-sectional shape strongly influences structural buildability. Layer pressing produces wider and flatter filaments, leading to less slender geometries with higher bending stiffness and increased resistance to buckling. Conversely, free-flow deposition generates more rounded filaments with reduced width and interlayer contact area, decreasing structural stiffness and promoting earlier instability.

    \item Similarly, in numerical simulations, the adopted cross-sectional representation can significantly affect buildability predictions. The influence is particularly relevant for free-flow deposition, where capturing the rounded filament geometry is essential, whereas layer-pressing strategies are less sensitive due to their flatter cross-sections, which are better represented by simple rectangular approximations.

    \item A useful parameter for assessing the approximation error in rectilinear walls is the \emph{roundness ratio} ($w_1/w_c$), defined as the ratio between the maximum filament width $w_1$ and the interlayer contact length $w_c$. This parameter provides a compact measure of the deviation of the filament cross-section from an ideal rectangular shape and was found to correlate strongly with the magnitude of the approximation error.
\end{itemize}

\begin{figure}[t]
    \centering
    \includegraphics[width=0.9\linewidth]{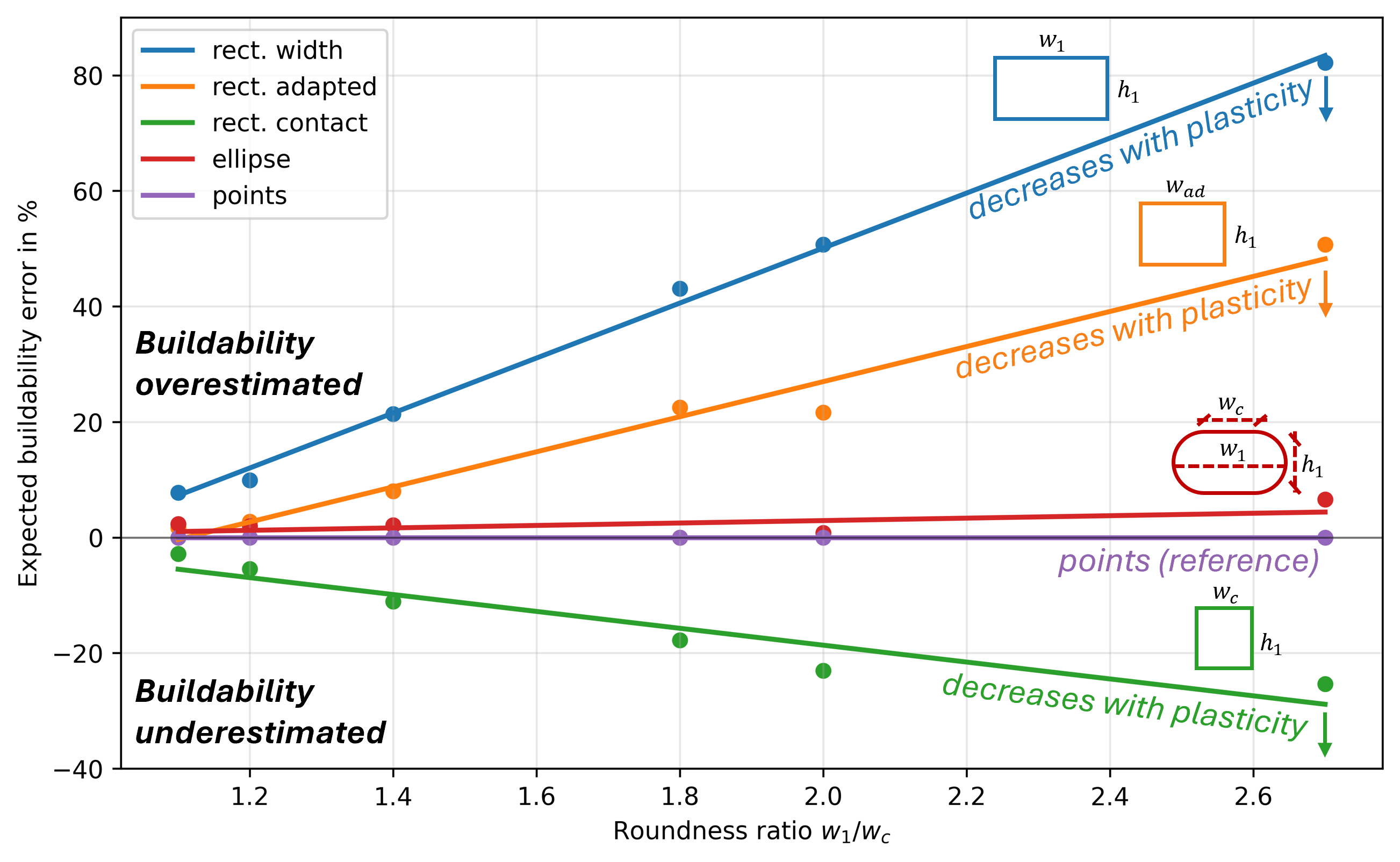}
    \caption{Simplified design chart indicating buildability percentage errors associated to different filament cross-sections discretization choices at varying of the roundness ratio for rectilinear walls.}
    \label{fig:design_chart}
\end{figure}

Based on these observations, a quantitative design chart is developed to estimate the error associated with buildability predictions obtained using different filament cross-sectional representations strategies for the case of perfectly elastic rectilinear walls. Figure~\ref{fig:design_chart} presents the proposed design chart, which relates the expected prediction error to the filament roundness ratio. The chart was derived from the results reported in Figure~17 by fitting linear regression trends to the data obtained for each cross-sectional approximation.

Several observations can be drawn from Figure~\ref{fig:design_chart}. First, the linear regressions provide a good representation of the numerical data, suggesting that the relationship between discretization error and roundness ratio can be reasonably approximated by a linear trend within the investigated range. Second, the figure highlights systematic biases associated with the different geometric approximations. The \textit{rect. width} and \textit{rect. adapted} filament representations tend to overestimate buildability, predicting higher number of layers than those obtained with the reference filament geometry. In contrast, the \textit{rect. contact} approximation systematically underestimates buildability. Finally, the \textit{ellipse} approximation shows the best overall agreement with the reference results (\textit{points}). Moreover, based on the previous parametric study (Table~\ref{tab:actlayers_youngs_modulus_relative}), the proposed chart is expected to be largely independent of the elastic modulus, as it represents a relative error measure. When plasticity effects are introduced, the analyses in Sect. \ref{sec:walls} observed a similar trend, although the magnitude of the error is affected: overestimating approaches become less conservative, while the \textit{rect. contact} approximation, which underestimates buildability, shows increased deviations. These effects of plasticity in the buildability error are indicated in the design chart by arrows.


Overall, the proposed design chart provides a practical framework for selecting filament representation strategies and quantifying the uncertainty associated with geometric simplifications in buildability simulations of thin-walled 3D-printed structures. Although it was derived for rectilinear walls, the present results demonstrate that the chart can also provide some meaningful guidance for curvilinear objects. Across the range of roundness ratios investigated, the discrepancies observed for both cylindrical configurations remained in fact within the uncertainty bounds predicted by the design chart. Specifically, the cylindrical case studies reveal that a curvilinear toolpath can enhance the geometric stiffness of the structure and reduce the sensitivity of buildability predictions to the adopted filament representation. As a result, although further research is needed to develop correction factors for highly curved structures, the design chart can in principle be expected to provide conservative error estimates when applied to curvilinear geometries.


The cylindrical-wall examples, which combine relatively large structural dimensions with a high number of buildable layers before collapse, also raise the important topic of computational cost. In all studies presented in this work, the element size was kept constant across the different filament representations to ensure a fair comparison of the resulting buildability predictions. The adopted mesh resolution was selected to adequately capture the geometry of the \textit{ellipse} and \textit{points} representations, which require a finer discretization due to their curved boundaries. Consequently, the performed simulations do not provide a representative basis for comparing computational efficiency. Due to their simpler and more regular geometry, rectangle-based models can generally be discretized with coarser meshes while maintaining accurate structural predictions. Therefore, they can significantly reduce computational time and memory requirements in large-scale buildability simulations. 

\begin{table}[h]
\caption{Decision map of layer shape approximation. In case of e.g. cylinders radii matters, too.}
\label{tab:results}
\begin{tabular}{lrr}
\toprule
Requirement & Roundness ratio $w_1/w_c \approx 1$ & Roundness ratio $w_1/w_c > 1$ \\
\midrule
High accuracy & ellipse & ellipse \\
Fast computation & any rectangle & \textit{rect. contact} (conservative) \\
Risk of overestimation acceptable & \textit{rect. adapted} or \textit{rect. width} (with caution) & \textit{rect. adapted} (with caution) \\
\bottomrule
\end{tabular}
\end{table}

In conclusion, we summarize the main findings in Table~\ref{tab:results}, and provide practical guidelines for selecting an appropriate cross-sectional representation:

\begin{enumerate}
    \item A point-wise representation of the filament cross-section is generally unnecessary for accurate buildability predictions. Among the investigated approaches, the \textit{ellipse} approximation consistently provided results close to the reference geometry while requiring a simpler geometric description, representing the best compromise between accuracy and modelling effort.

    \item Rectangular approximations remain valuable when computational efficiency or conservative estimates are prioritized. The \textit{rect. contact} discretization consistently underestimates buildability and can therefore be used as a conservative lower-bound approach, whereas the \textit{rect. width} formulation tends to overestimate buildability and should be applied with caution in safety-critical assessments.

    \item The \textit{rect. adapted} discretization merits separate discussion. Although it also tends to overestimate buildability, its error remains considerably smaller than that of \textit{rec. width}. Moreover, despite its simplicity, its geometry is designed to preserve the volume of the reference (most realistic) representation, unlike the other discretizations. Consequently, the total weight is preserved without the need for artificial density adjustment. This is particularly advantageous for dynamic and contact simulations, where density adjustment may be undesirable, as it can alter inertia, natural frequencies, and contact-related responses.

\end{enumerate}

\section{Conclusion}\label{sec:5}

This work presented a geometrically informed workflow for buildability assessment in 3D concrete printing, integrating the deep-learning-based filament shape prediction framework \textit{ShapeGen3DCP} with a layer-activation finite element model. By enabling the automatic generation of geometry-aware numerical models directly from material and process parameters, the proposed methodology eliminates the need for preliminary filament characterization or computationally expensive fluid-dynamics simulations. Application of the framework demonstrated that filament geometry can substantially influence buildability predictions and that incorporating realistic, process-informed cross-sectional representations significantly improves the predictive capability of standard FEM-based approaches.

The validation and parametric studies demonstrated that the influence of filament geometry on buildability predictions depends strongly on both the adopted printing strategy and the governing failure mechanism of the printed structure. In particular, rounder filaments associated with free-flow deposition were found to be substantially more sensitive to geometric simplifications than the flatter filaments produced by layer-pressing strategies. Likewise, rectilinear walls exhibited greater sensitivity than geometrically stiffer configurations, such as cylindrical structures. In contrast, the influence of filament geometry was found to be only weakly dependent on the constitutive material behaviour, with limited sensitivity to variations in elastic stiffness or to the adoption of elastic versus elasto-plastic material models. 

Among the investigated filament cross-sectional representations, elliptical geometries consistently provided the best compromise between prediction accuracy and modelling simplicity. The use of a more complex realistic polyline representation resulted in only marginal improvements in predictive accuracy, while introducing substantially greater geometrical complexity. On the other side, when rectangular representations are preferred due to their regular mesh topology and computational efficiency, enforcing volume conservation through an appropriate dimensioning strategy can significantly enhance the reliability of the resulting simulations.

These findings motivate the development of a new class of \emph{geometrically informed} buildability models, in which extrusion process parameters and the resulting filament geometry are explicitly integrated into structural simulations. In this work, this paradigm was realized by coupling \textit{ShapeGen3DCP} with a layer-activation FEM framework, demonstrating that machine-learning-based geometry prediction can be effectively integrated into existing numerical workflows without excessive increasing computational complexity. More broadly, the proposed framework illustrates how data-driven geometry prediction can bridge the gap between process modelling and structural analysis, enabling more realistic yet computationally efficient simulations of the printing process.

From a practical perspective, the proposed workflow and the accompanying design guidelines provide a straightforward methodology for selecting an appropriate geometric representation according to the required balance between accuracy and computational cost. The presented design chart allows users to estimate the uncertainty introduced by different filament cross-sectional representations based on the filament roundness ratio. The proposed framework therefore represents a practical step towards more reliable and predictive numerical tools for the design, optimization, and structural assessment of 3D concrete printing processes.

\section*{Acknowledgements}
The authors acknowledge the Gastwissenschaftlerprogramm of Bundesanstalt für Materialforschung und -prüfung (BAM), which supported a two-month research stay of G. Rizzieri at BAM during which the work presented in this paper was initiated. Additionally, the first author acknowledges the project "Material and Process Modelling for Lunar 3D Printing", CUP D47G25000060006, pursuant to Notice No. 47 of 20/02/2025 (‘Decree for the recruitment of international post-doctoral researchers’), PNRR, Mission 4, Component 2, Investment 1.2, financed by the European Union - NextGeneration EU.

\section*{Data availability}
The workflow implementation, including all scripts and results, is available as a dataset publication in \cite{rizzieri2026a}.

\bibliographystyle{elsarticle-num.bst}
\bibliography{Bibliography}

\end{document}